%% file: Multipole_RVM.tex
\documentclass[twocolumn,appendixfloats]{openjournal}

\usepackage{comment}
\usepackage{graphicx}
\usepackage{natbib}
\usepackage{xcolor}
\usepackage{amsmath}
\usepackage{amssymb}
\usepackage{txfonts}
\usepackage{gensymb}
\usepackage[
    colorlinks=true,
    linkcolor=violet,      % internal references: sections, figures, equations
    citecolor=teal,      % bibliography citations
    urlcolor=violet     % web links
]{hyperref}

\input{macros.tex}

\newcommand{\vecBD}{\textbf{B}_\mathrm{D}}
\newcommand{\vecBQ}{\textbf{B}_\mathrm{Q}}
\newcommand{\BD}{B_\mathrm{D}}
\newcommand{\BQ}{B_\mathrm{Q}}
\newcommand{\fQ}{f_\mathrm{Q}}
\newcommand{\RNS}{R_\mathrm{NS}}
\newcommand{\RLC}{R_\mathrm{LC}}
\newcommand{\thD}{\theta_\mathrm{D}}

\newcommand{\thobs}{\theta_\mathrm{obs}}
\newcommand{\Po}{$P=1$\,ms}
\newcommand{\Pf}{$P=5$\,ms}
\newcommand{\PAdiff}{\sigma_{\mathrm{PA_{D}}}}
\newcommand{\PAmnrms}{\sigma_{\mathrm{PA}}}
\newcommand{\PAslice}{\sigma'_{\mathrm{PA_{D}}}}

\graphicspath{{./}}

\shorttitle{RVM in multipole magnetospheres}

\begin{document}
\journalinfo{The Open Journal of Astrophysics}
%\submitted{submitted xxx, revised yyy, accepted zzz}

\title{Rotating Vector Model predictions in millisecond pulsar magnetospheres with dipole and quadrupole field components}

%% use optional labels to link authors explicitly to addresses:
%% \author[label1,label2]{}
%% \address[label1]{}
%% \address[label2]{}

\author{
A. V. Bilous$^{1}$
}
\email{hanna.bilous@gmail.com}

\affiliation{$^1$Independent researcher}

\begin{abstract}
The rotating vector model (RVM) is widely used to interpret pulsar radio polarization 
under the assumption of a dipolar magnetic field. Motivated by tentative evidence for 
non-dipolar magnetic structure in millisecond pulsars provided by recent modeling of 
surface thermal X-ray pulsations, we examine how a global 
quadrupole component modifies position-angle (PA) behavior within the classical RVM 
framework. We numerically integrate magnetic field lines in static vacuum 
magnetospheres composed of a centered dipole component that is inclined with respect to 
the spin axis, and a centered quadrupole component aligned with it. We then compute the open 
field line regions (OFLRs), emission skymaps, and PA curves 
for a range of dipole inclinations, relative quadrupole strengths, and light cylinder radii
corresponding to two trial spin periods of 1\,ms and 5\,ms.
We find that quadrupole-modified magnetospheres produce a wide variety of PA 
behaviors and substantially alter the morphology and total area of OFLRs, 
with the OFLR area varying non-monotonically with quadrupole strength and, 
in some cases, exceeding that of the corresponding pure dipole. 
Depending sensitively on the field configuration and viewing geometry, three broad PA 
regimes can be distinguished: a) nearly dipolar PA swings that are largely insensitive 
to emission altitude; b) moderately distorted, altitude-dependent PA tracks that can 
bias dipolar RVM fits; c) strongly dispersed PA behavior associated with emission 
from different OFLRs, which can remain incompatible with a single dipolar 
PA track even within narrow emission altitude ranges.
These results demonstrate that dipole-like PA curves do not uniquely imply 
a purely dipolar magnetic field, while global quadrupole components alone 
are insufficient to reproduce the full complexity of observed millisecond pulsar PA profiles.
\end{abstract}

\section{Introduction}

Recent modeling of surface thermal X-ray pulsations from seven nearby
recycled pulsars has revealed emitting-region geometries that sometimes
are difficult to reconcile with the simplistic picture 
\citep[see][for the most recent work on each pulsar and references therein]{Kini2026,Dittmann2024,Miller2026,Qi2025,Mauviard2025,GonzalesCaniulef2026,Mauviard2026}. 
In this picture, thermal emission is 
induced by particles accelerating downward in the open field line regions (OFLRs)
around the magnetic poles of an oblique dipole magnetosphere.

One of the early suggestions to 
explain the discrepancy between observed and theoretical locations of heated regions 
was to introduce more complex global 
magnetic field configurations that include higher-order multipole 
components \citep{Chen2020,Kalapotharakos2021}. The question of whether
such components are indeed required to produce heated regions outside 
the conventional dipolar polar caps is still open \citep{Carrasco2023}. 
However, the presence of global multipole components is not strictly prohibited 
from a theoretical standpoint. For recycled pulsars, magnetic field 
multipolarity is naturally expected in the burial scenario associated 
with the accretion process \citep{Suvorov2020}.

\begin{figure*}
\centering
 \includegraphics[width=0.95\textwidth, clip]{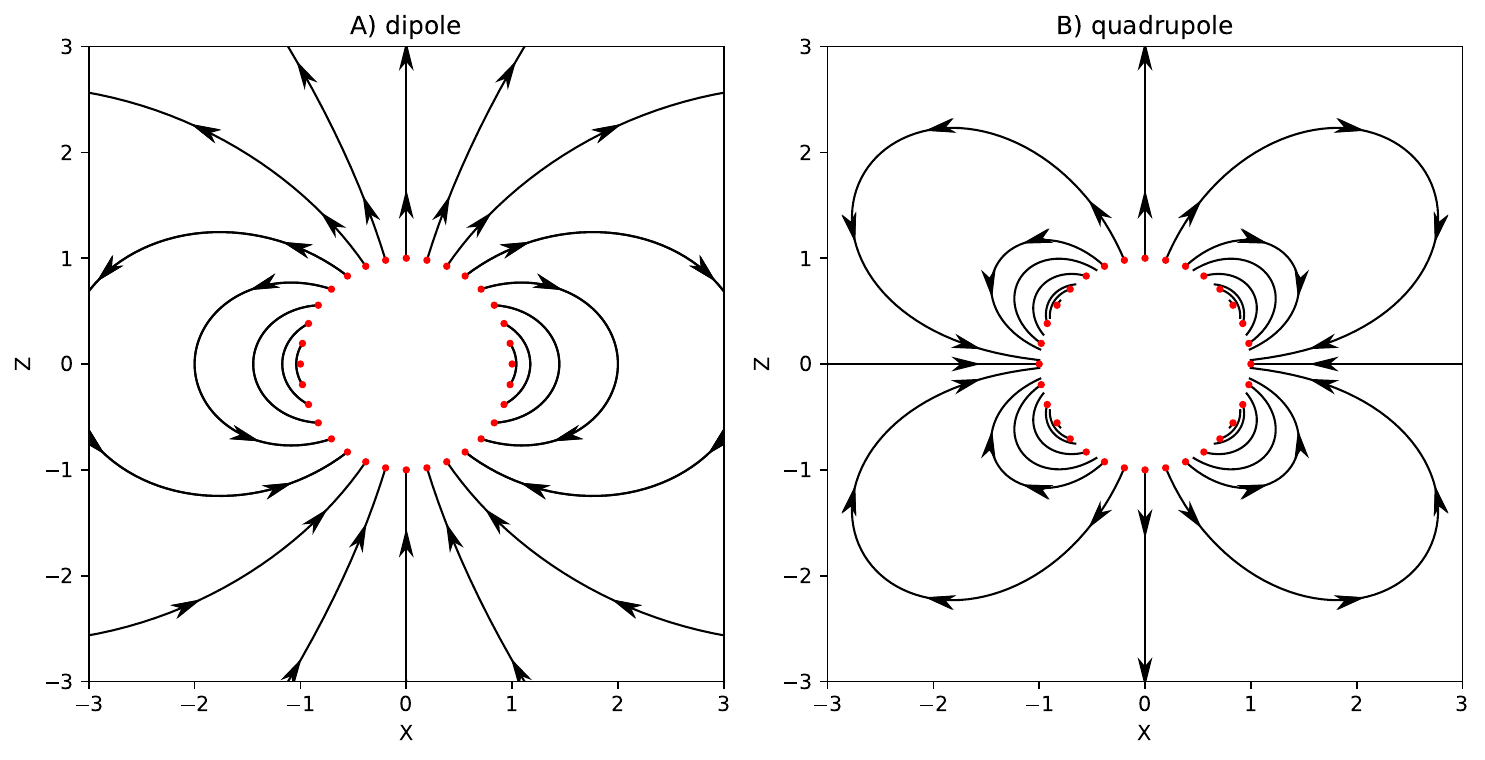}  
 \caption{Field lines for the dipole (left) and quadrupole (right) magnetic fields, 
 with the magnetic moment aligned with the Z axis. Only the YZ plane is shown, 
 exploiting the rotational symmetry about the Z axis. Field lines are 
 integrated outward from surface footpoints (red dots).
}
\label{fig:prime_meridian_lines}
\end{figure*}

The deviations from a centered oblique dipole field have been 
invoked by pair-cascade theorists to obtain a pair multiplicity 
large enough to produce radio emission \citep{Harding2011,Gil2002}.
Yet, at the altitudes of radio emission generation, the global 
magnetic field is widely assumed to be dipolar.
A key observational argument in favor of such a simple 
configuration comes from the linearly polarized component of pulsed 
radio emission. For some pulsars, the position angle (PA) of the linearly 
polarized radio emission varies according to a characteristic S-shaped 
curve as the star rotates past the observer. Although the radio 
emission mechanism remains poorly constrained \citep{Beskin2018}, 
the observed S-shaped swing can be explained in a purely geometric 
manner by assuming that both the direction of the emission beam and the plane of 
linear polarization are determined by the orientation of the magnetic 
field \citep[the so-called rotating vector model, RVM,][]{Radhakrishnan1969}. 
The resulting classical analytical 
prescription for the position angle in the field of an oblique dipole 
has since formed the basis of geometric constraints for many sources, 
enabling estimates of the inclination angle of the dipole magnetic 
moment and the observer’s impact angle.

On the other hand, there has always been evidence for   
far more complex PA behavior than could be described by RVM. 
Among non-recycled pulsars (with long periods 
and large light cylinder radii), approximately half of the sources 
in large-scale studies cannot be satisfactorily described by 
the RVM prescription \citep{Johnston2023,Mitra2023,Wang2023}. 
Disentangling the contributions of orthogonal polarization modes 
and the propagation of radio waves through the magnetospheric 
plasma can yield improved fits within an oblique dipole field 
framework \citep{Beskin2012,Johnston2024}.

For recycled millisecond pulsars (MSPs), PA variations across the pulse 
profile are particularly complex, often frequency-dependent, 
and generally cannot be fitted with the oblique dipole RVM \citep{Dai2015,Wahl2022}. 
In the largest published sample of MSPs to date (56 sources), only 
five were found to be consistent with the RVM description \citep{Xu2025}.

To date, there is no clear answer to the question of whether the PA
of linearly polarized radio emission can be used to constrain the
presence of global multipole components in pulsar magnetospheres.
Existing studies have considered a limited set of specific field
configurations, including an aligned dipole combined with a slightly
oblique quadrupole \citep{Qiu2023}, or multipole configurations
derived from observations of PSR~J0030+0451 \citep{Sur2024}. To the best
of our knowledge, no systematic investigation has been carried out to assess
how a multipole component affects the classical PA swing
of an arbitrarily oblique dipole.

In this study, we introduce a minimal modification to the classical centered
oblique dipole model by adding a centered quadrupole component 
aligned with the spin axis. This choice keeps the
parameter space sufficiently small for systematic exploration.
However, the resulting configurations are not intended to represent
the full complexity of realistic magnetic fields, which may include
displaced components or contributions from multiple multipole orders
\citep{Petri2015,Kundu2026}.

In the following sections we numerically compute the field lines of 
static vacuum magnetospheres and explore the morphology of OFLRs and 
the behavior of PA curves for a range of dipole inclination angles and 
relative quadrupole strengths. We do not model relativistic effects, 
wave propagation through the magnetospheric plasma, or changes in the 
field morphology due to magnetospheric currents. The resulting PA curves 
can be compared directly with the dipolar RVM prescription and used to 
assess how the quadrupole component modifies the classical S-shaped PA swing.

\section{Magnetic field geometry and RVM framework for pure dipole and quadrupole fields}
In this work, we perform numerical integrations of magnetic field lines for 
a magnetic field consisting of a centered oblique dipole component (D)
and a centered axisymmetric (azimuthal order $m=0$) quadrupole component (Q): 
\begin{equation}
 \mathbf{B} = \mathbf{B}_{\rm D} + \mathbf{B}_{\rm Q}.
 \label{eq:total_field}
\end{equation}

Before exploring the results of the numerical integrations, it is useful 
to establish intuition by reviewing the shapes of the field lines, the 
locations of the OFLRs, and the PA behavior for purely 
dipolar and purely quadrupolar field configurations.

In a spherical magnetic coordinate system, %set by vectors $[\textbf{e}_r,\ \textbf{e}_\theta,\ \textbf{e}_\phi]$
the field components have the following form \citep{Kalapotharakos2021}:
\begin{equation}
\begin{aligned}
\vecBD & = \BD\left(\frac{\RNS}{r} \right)^3\left[ \cos\theta,\ \frac{1}{2}\sin\theta,\ 0 \right],  \\
\vecBQ & = \BQ\left(\frac{\RNS}{r} \right)^4\left[ \frac{1}{2}(3\cos^2\theta-1),\ \cos\theta\sin\theta,\ 0 \right].
\end{aligned}
\label{eq:comp_field}
 \end{equation}
Here $r$ is the radial distance to the center of the star, $\theta$ is colatitude, and $\RNS$ is the stellar radius.

The equation for a magnetic field line follows from the condition that the tangent to the line is parallel to 
the local magnetic field:
\begin{equation}
 d\textbf{r} \times \textbf{B} = 0.
\end{equation}

For pure dipole and quadrupole fields, the field line equation has a simple analytical 
form in the magnetic spherical coordinate system:
\begin{equation}
 r = \RNS \dfrac{\sin^2\theta}{\sin^2\theta_0}
\end{equation}
for the dipole field, and 
\begin{equation}
 r = \RNS \dfrac{\sin\theta}{\sin\theta_0}\sqrt{\dfrac{\cos\theta}{\cos\theta_0}}
\end{equation}
for the quadrupole field. Here, $\theta_0$ is the colatitude of the field line footpoint
at $r=\RNS$. 

\begin{figure*}
\centering
 \includegraphics[width=0.9\textwidth, clip]{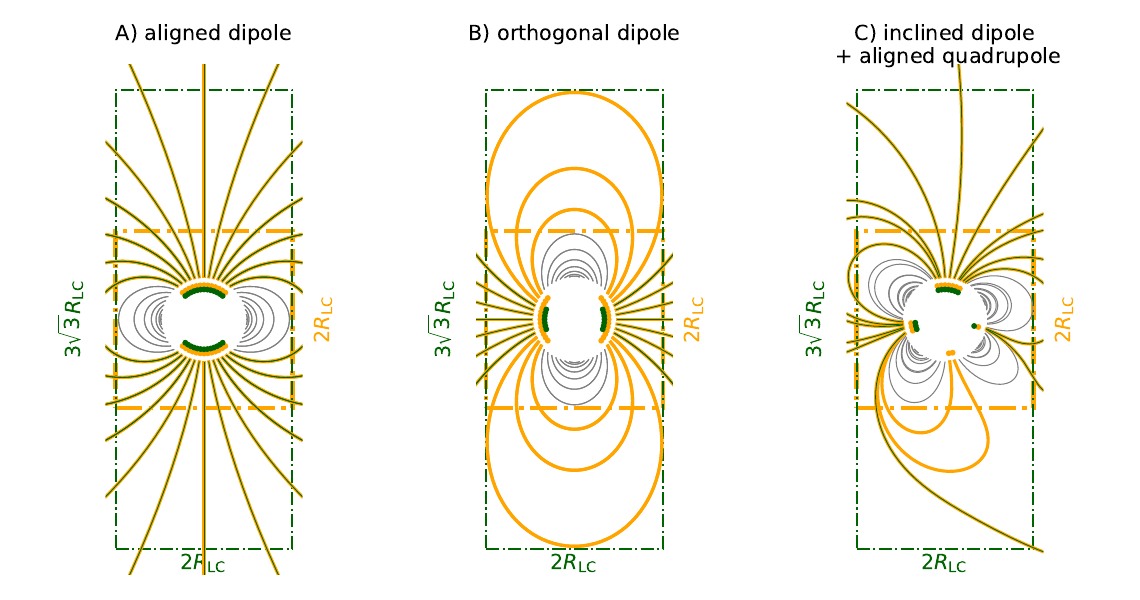}
   \caption{Illustration of OFLR size dependence on different conventions 
   about the height of the LC. In all three subplots, two
   choices of LC heights are shown with dot-dashed lines: $h_\mathrm{LC} = 2\RLC$ in orange,
   and $3\sqrt{3}\RLC$ (green). The same color conventions apply to 
   field lines that penetrate each light cylinder. Thin black lines mark
   closed field lines. OFLRs are marked with curves inside 
   the star. \textit{Left:} aligned dipole field. OFLRs do not 
   depend on the assumed LC height. \textit{Middle:}
   orthogonal dipole field. For $h_\mathrm{LC}>3\sqrt{3}\RLC$ 
   the size of the OFLR does not depend on $h_\mathrm{LC}$. \textit{Right:}
   a superposition of oblique dipole and aligned quadrupole fields with the adopted 
   choice of LCs. The shape and number of disjoint OFLRs depend on the choice 
   of $h_\mathrm{LC}$.}
\label{fig:OFL_vs_LC_height}
\end{figure*}

Field lines of both dipole and quadrupole components are shown 
in Fig.~\ref{fig:prime_meridian_lines}. Note that field lines of the dipolar 
field are symmetric with respect to the equatorial plane:
a field line starting at the colatitude $\theta_0$ is the same as the one starting at the 
colatitude $180\degree-\theta_0$. For the quadrupole field, no such symmetry exists: lines starting 
at $\theta_0$ will return to the star surface at angles different from $90\degree - \theta_0$.

Open field lines are defined as lines that penetrate the light cylinder (LC), 
a surface defined by the condition that corotation velocity is equal to 
the speed of light $c$. The radius of the light cylinder, $\RLC$, can be calculated from the spin 
period $P$ with the following equation:
 \begin{equation}
 \RLC = \frac{cP}{2\pi}\simeq 4.77 \times 10^4\ \mathrm{km} \left(\frac{P}{\mathrm{s}}\right) \simeq 4.77 \times \RNS \left(\frac{P}{\mathrm{ms}}\right).
\end{equation}

If the magnetic moment is parallel or antiparallel to the spin axis, then 
the field has rotational symmetry around the spin axis.
Open field lines come from circular regions around the north and south 
poles, and, for the quadrupole field, also from an equatorial strip. 
The limiting magnetic colatitudes of OFLRs can be estimated from 
$d(r \sin\theta_\mathrm{LC}) = 0$. For the dipole field this relation results in 
$\theta_\mathrm{LC} = 90\degree$ and $\sin\theta_\mathrm{OFLR} = \sqrt{\RNS/\RLC}$ 
for the north pole and $180\degree - \theta_\mathrm{OFLR}$ for the south pole.

For the quadrupole field, $\cos\theta_\mathrm{LC} = 5^{-0.5}$ 
and the polar OFLRs are bounded by $\theta_\mathrm{OFLR}\approx 34\degree(\RNS/\RLC)$ if we assume 
$\RNS/\RLC$ to be small enough to neglect the higher-order terms of the Taylor 
expansion of trigonometric functions. With the same accuracy, the meridional width 
of the equatorial strip is $41\degree(\RNS/\RLC)^2$. 

In the classical RVM the PA is determined by the projection 
of the field line binormal vector onto the observer’s plane of the sky. 
For both an oblique dipole and an oblique quadrupole field, the PA at a given 
spin longitude does not depend on the emission height and is 
described by a simple analytical expression:
\begin{equation}
\label{eq:RVM}
 \tan \mathrm{PA} =
 \dfrac{\sin\alpha\,\sin\phi}
 {\sin\thobs\cos\alpha - \cos\thobs\sin\alpha\cos\phi},
\end{equation}
where $\thobs$ is the colatitude of the observer, $\alpha$ is the angle 
between the spin axis and the magnetic moment, and $\phi$ is 
the spin longitude \citep{Radhakrishnan1969,Hibschman2001,Lyutikov2016,Qiu2023}.

The observed PA is defined in terms of the Stokes parameters $Q$ and $U$ as
$\mathrm{PA} = 0.5\arctan(U/Q)$, and is therefore restricted to the range
$[-90\degree, 90\degree]$. Accordingly, the PA given by Eq.~\ref{eq:RVM}
must be wrapped into this interval.

\section{Numerical setup and explored parameter space}
\label{subsec:simulation}

\begin{figure*}
\centering
 \includegraphics[width=\textwidth]{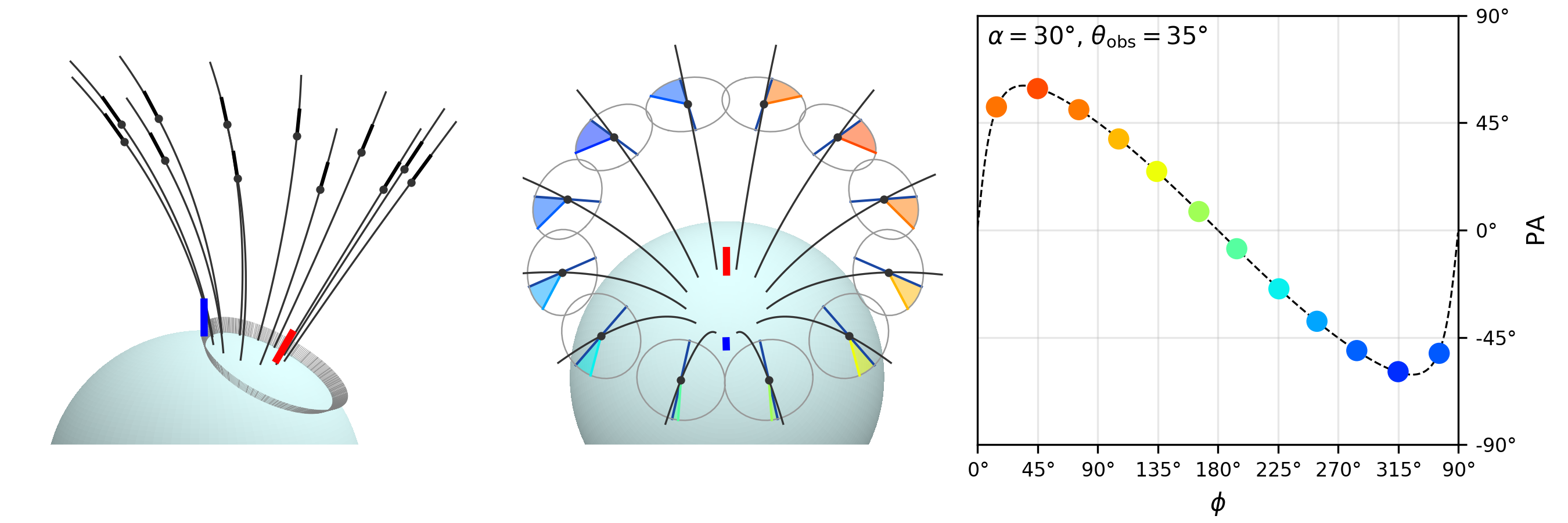}
 \caption{Schematic illustration of the RVM for an oblique dipole rotator with 
 inclination angle $\alpha = 30\degree$ and observer's colatitude $\thobs = 35\degree$. 
 Emission is sampled from points uniformly spaced in observed spin longitude $\phi$ (Eq.~\ref{eq:emission_point})
 and located at constant altitude above the star.
 \textit{Left} and \textit{middle} panels show a 3D view of field lines with emission points emerging from 
 the stellar surface (pale blue). Spin and magnetic dipole axes are indicated by blue and red arrows, respectively. 
 Additionally, the \textit{left} panel shows the last open field lines for \Po\ trimmed very close 
 to the stellar surface, together with the direction of the LOS at each emission point (black line). 
 The \textit{middle} panel adds a plane perpendicular to the LOS 
 at the emission point (gray circle). The projection of spin axis 
 onto this plane, $\mathbf{\Omega}_p$, is shown by a dark blue line, 
 while the curvature vector $\mathbf{n}$ is shown by a thick colored line. 
 The angle between $\mathbf{\Omega}_p$ and $\mathbf{n}$ is colored according to its value (Eq.~\ref{eq:PA_numeric}).
 \textit{Right:} PA as a function of spin longitude $\phi$ for the same 
 configuration. Marker colors correspond to PA values. The dashed black curve shows the analytical RVM prediction given by Eq.~\ref{eq:RVM}. 
}
\label{fig:RVM_cartoon}
\end{figure*}

We integrate field lines in the 
corotating frame, with the dipole moment confined to the $XZ$ 
plane. In this setup, our magnetic field configuration is fully
specified by two parameters: the colatitude of the dipole moment, 
$\thD$, and the relative strength of the quadrupole 
component, $\fQ \equiv B_{\rm Q}/B_{\rm D}$. The extent of the 
OFLR is then determined by the LC 
radius corresponding to the chosen $P$.

We explore all combinations of $\thD$ in the range 
$[\thDmin,\,\thDmax]$ with 
a step size of $\thDstep$.
The choice of $\fQ$ and 
$P$ is primarily constrained by the resolution of the footpoint 
grid and the associated computational cost. After an initial 
exploration, we restricted $\fQ$ to the range 
$[\BQmin,\,\BQmax]$, with a step size of 
$\BQstep$. This range is comparable to that inferred 
by \citet{Kalapotharakos2021} from multiwavelength pulse profile 
modeling of PSR~J0030+0451, although some of their models 
required values as high as $\fQ\simeq 8.7$.

We considered two representative spin periods, \Po\ (close to 
the fastest known MSP, PSR~J1748$-$2446ad) and \Pf.
Together, these values approximately span the range of 
spin periods of NICER pulsars for which pulse profile modeling is 
currently available: from $P=2.9$\,ms of PSR~J0740+6620 to 
$P=5.8$\,ms of PSR~J0437$-$4715.

For oblique rotators, both dipolar and quadrupolar, the extent of 
the OFLR on the surface of the star depends on the assumed height of the 
light cylinder. This dependence is illustrated in Fig.~\ref{fig:OFL_vs_LC_height}. 
For the subsequent analysis, we adopt the following light cylinder height:
\begin{equation}
h_{\rm LC} = 3\sqrt{3}\,R_{\rm LC}. 
\label{eq:LC}
\end{equation}
For an orthogonal dipole field, this value corresponds to twice 
the maximum distance along the $Z$ axis reached by the field line 
that attains a cylindrical radius of $R_{\rm LC}$. This choice 
is admittedly simplistic, but is consistent with the other 
 assumptions adopted in this study. In reality, the topology 
of magnetic field lines and, consequently, the shape of the OFLRs 
are modified by magnetospheric currents \citep{Bai2010,Cerutti2017}. 

\subsection{Field line integration}
All numerical calculations were performed using the custom software suite 
\texttt{psr-magpy}\footnote{\url{https://github.com/AnnaBilous/psr-magpy}}.
Field lines were integrated outward starting from a grid of footpoints 
on the stellar surface. For footpoints at which the surface magnetic 
field is directed toward the star, the direction of the field vector 
was inverted for the purposes of numerical integration. The footpoints 
were distributed using a \texttt{HEALPix} grid (Hierarchical Equal Area 
isoLatitude Pixelization\footnote{\url{http://healpix.sf.net}}, implemented via the \texttt{healpy} 
Python package \citep{Gorski2005,Zonca2019}. Originally developed for the analysis of cosmic 
microwave background data, the \texttt{HEALPix} scheme subdivides the spherical 
surface into pixels of equal area, with pixel centers arranged in rings of constant latitude.

To perform an initial mapping of the locations of the OFLRs, we 
employed a \texttt{HEALPix} grid with a characteristic parameter 
$\mathtt{nside}=\nsideInitial$, resulting in $\nPixInitial$ pixels, each covering 
approximately $\pixAreaDegSqInitial~\mathrm{deg}^2$ on the stellar surface. Field 
lines were integrated outward up to a radial distance of 
$R_{\rm lim}=\Rmax\,\RNS$ or until they returned to the 
surface of the star. 

To improve computational efficiency, 
three approaches were used: the calculations were parallelized and 
executed in batches; numerical integration was performed using 
the \texttt{\solverBackend}\footnote{\url{https://github.com/jrenaud90/CyRK}} 
package \citep{Renaud2025}, 
which provides a high-performance 
Cython-based solver similar to \texttt{scipy.integrate.solve\_ivp}; 
and magnetic field computations used 
\texttt{\fieldBackend}-compiled functions.
The resulting field line trajectories were sampled at uniform arc length 
intervals of $\dt\,R_{\rm NS}$ and stored on disk. 

For the spin periods of \Po\ and \Pf, the light cylinder radii are 
\RLCedgeO\ and \RLCedgeF. At 
the initial \texttt{HEALPix} resolution, the equatorial strip associated 
with the quadrupole component is not well resolved. Increasing the 
resolution over the entire stellar surface, however, is computationally 
expensive. We therefore identified the OFLRs for the $P=\Pmin$\,ms 
case and recomputed the field lines in those regions using a refined 
grid with $\mathtt{nside}=\nsideRefined$, corresponding to a factor of \nsideRefinementFactor\ 
improvement in spatial resolution.

We verified the accuracy of the numerical integration by tracing each
field line back from its termination point (either at $\RNS$ or at
$R_{\rm lim}$) and comparing the coordinates of the resulting
termination point with those of the original surface footpoint. For almost all of
the roughly one million lines integrated per experiment, the separation between these two
points was very small, typically of order $10^{-6}\RNS$.
Here, an experiment is defined as a set of field lines integrated for a specific
field configuration and spin period.

Eleven experiments contained lines with separation distances of $10^{-3}\RNS$. These lines
were excluded from all subsequent analyses.
Of these, eight experiments with $\thD<90\degree$
contained 2$-$4 such lines per set, including
two experiments with $\fQ=0$. The remaining three experiments, all with
$\thD=90\degree$ and $\fQ>1.5$, had about 1000 runaway lines
each, with all such lines originating at the equator. In this region,
the dipole and quadrupole field components nearly cancel
each other out for $\thD=90\degree$, leading to numerical integration failure.

\subsection{Emission points selection}

Emission points were determined as follows. For each field line, 
we computed the tangent vector $\mathbf{t}$ and evaluated 
its colatitude $\theta_t$ and azimuth $\phi_t$ in the corotating 
frame. Emission points corresponding to the observed spin longitude 
$\phi$ and viewing angle $\thobs$ were identified by two requirements:
\begin{equation}
\label{eq:emission_point}
\begin{aligned}
|\theta_t - \thobs| < \delta_\theta, \\
\phi = 360\degree - \phi_t.
\end{aligned}
\end{equation}
The tolerance around $\thobs$, $\delta_\theta$, represents the finite width of the emission beam 
produced by individual relativistic particles \citep{Gangadhara2004}. 
Reversing the $\phi$ coordinate reflects our convention that the pulsar 
rotates counterclockwise as seen by an observer with $\thobs<90\degree$; 
consequently, in the corotating frame the line-of-sight (LOS) moves clockwise in azimuth.

Further, we imposed constraints on the emission height. For the lower limit, 
we took the first sampled point on the field line that did not lie on the stellar surface.
This corresponded to $r_\mathrm{min} = \rMinEmiss\RNS$, which roughly matches
the altitude of the first pairs produced by inverse Compton photons for MSPs \citep{Harding2002}. 
For the upper limit, we restricted emitting points to
radial distances below $r_\mathrm{max} = \rTrimCoeff\RLC$ from the center of the star,
broadly within the range of estimated emission heights for MSPs \citep{Gangadhara2006,Petri2023}.

The PA at the emission point was computed as the angle between the
field line curvature vector $\mathbf{n}$ and the projection of the spin vector onto the plane
perpendicular to the tangent direction $\mathbf{t}$,
$\mathbf{\Omega}_{p} = \mathbf{\Omega} - (\mathbf{\Omega}\cdot\mathbf{t})\cdot\mathbf{t}$
\citep{Qiu2023,Hibschman2001}:
\begin{equation}
\label{eq:PA_numeric}
\tan \mathrm{PA} = \dfrac{\mathbf{\Omega}_{p}\cdot(\mathbf{t}\times\mathbf{n})}{\mathbf{\Omega}_{p}\cdot\mathbf{n}}.
\end{equation}
Figure~\ref{fig:RVM_cartoon} illustrates the vectors involved in the PA calculation for a simple case of an oblique dipole field.

The calculated PA was wrapped to the interval $[-90\degree,\,90\degree]$, corresponding to the
smaller signed angle measured toward the projected spin axis $\mathbf{\Omega}_{p}$. 
Points affected by numerical noise due to very large curvature radii were excluded from the analysis.

\section{Open field line regions in magnetospheres with dipole and quadrupole components}
\label{sect:OFLR}

\begin{figure}
\centering
 \includegraphics[width=0.9\columnwidth, clip]{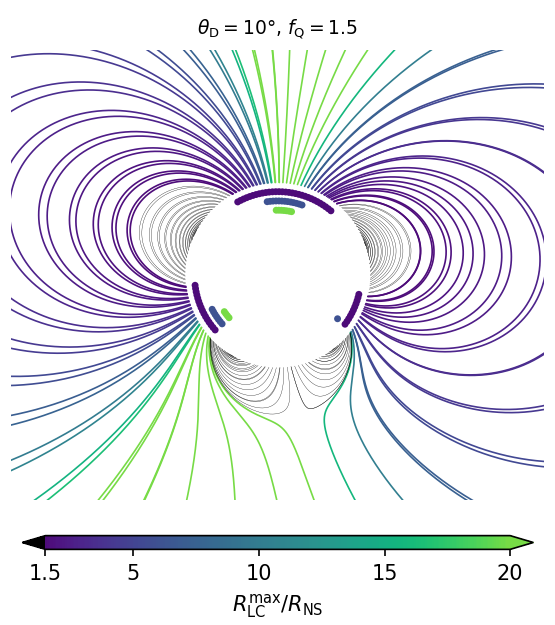}
   \caption{Illustration of two qualitatively different behaviors of OFLR boundaries for 
   a multipole field with $\thD=\thDedgetype$ and $\fQ=\fQedgetype$. Field lines are colored according 
   to the maximum $\RLC$ for which they remain open. Colored segments on the stellar 
   surface show OFLRs for $\RLC$ = $\RLCedgemarks$. 
   The northern OFLR and the northern boundary of the southern OFLR contract gradually 
   with increasing $\RLC$, reflecting a smooth change of the maximum extent of the neighboring
   field lines. By contrast, the southern boundary of the latter remains 
   at precisely the same location over the full range of $\RLC$ shown, indicating a sharp 
   transition in the maximum extent of neighboring field lines.}
\label{fig:OFLR_edge_type}
\end{figure}

In this section we provide a general overview of the
3D field line geometry together with the surface 
morphology of OFLR footpoints. Several important remarks should 
be made first. Visualizing 3D field lines of a 
complex magnetic field on a flat image plane is inherently 
challenging and necessarily involves simplifications.
Plotting field lines originating from all \texttt{HEALPix} footpoints is
impractical given the high density of the footpoint grid, while
randomly selected subsets may obscure the underlying bulk field line
geometry. Projecting field lines onto a plane 
(Fig.~\ref{fig:OFLR_edge_type}) can reveal some aspects of their
structure, but does not capture the full 3D geometry.

\begin{figure*}
\centering
 \includegraphics[width=\textwidth, clip]{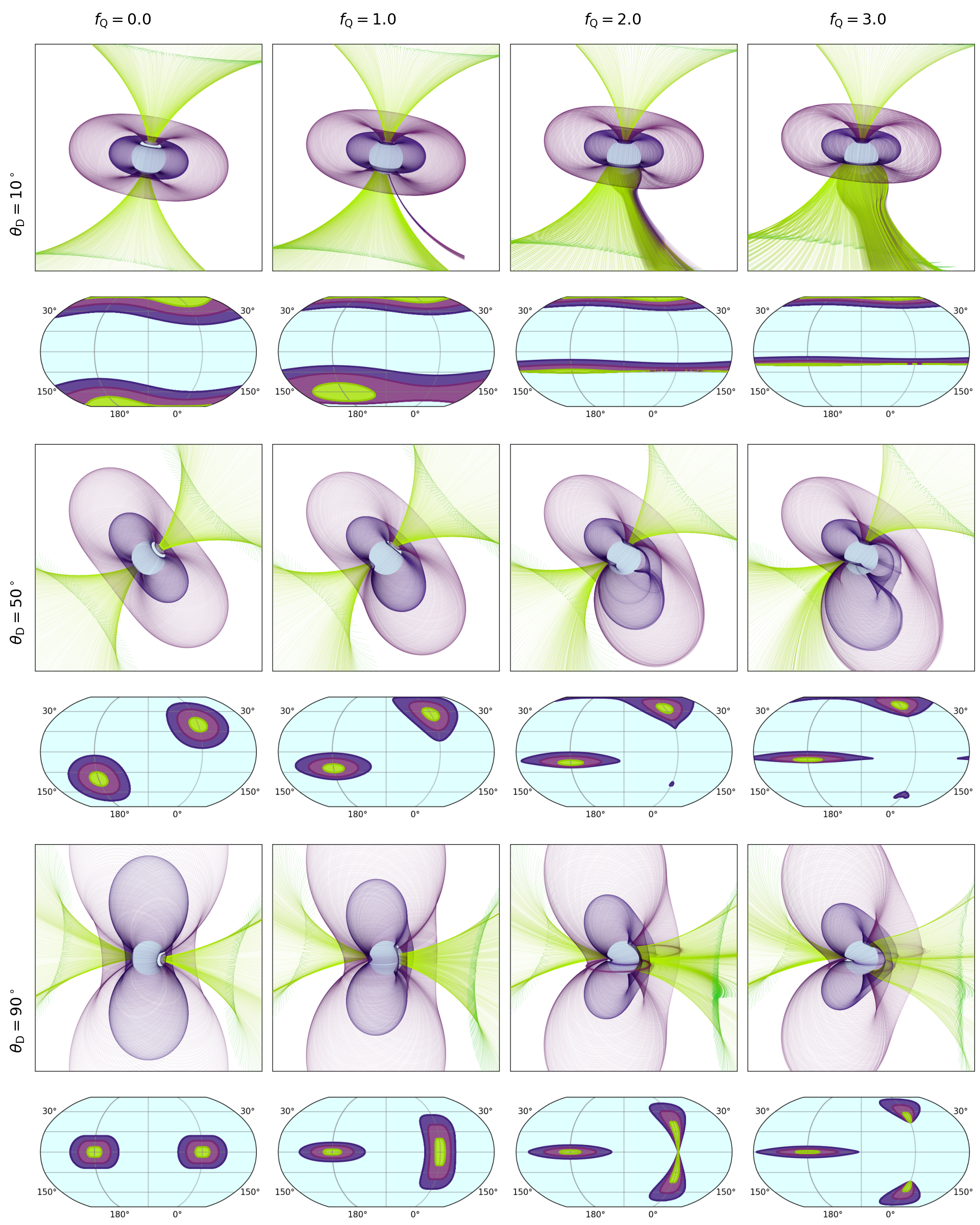}
   \caption{Last open magnetic field lines and open field line regions for spin periods 
   of \PillA\ (dark purple), \PillB\ (light purple), and \PillC\ (green), shown for 
   $\thD = \thDillustr$ (rows) and $\fQ = \fQillustr$ 
   (columns). In each panel, the upper subpanel shows last open field lines, 
   while the lower subpanel shows the corresponding OFLR
   footpoints on the stellar surface. Lines are plotted only to the extent of their 
   light cylinders with the height defined by Eq.~\ref{eq:LC}.
   All OFLR maps are plotted with $\phi = 90\degree$ at the center, to minimize
   projection distortion at the map edges.}
\label{fig:FL_overview}
\end{figure*}

To address these limitations, we chose
to plot the last open magnetic field lines for a set of selected
spin periods in 3D space (Fig.~\ref{fig:FL_overview}).
This representation provides a more direct view of
variations in the large-scale field line geometry while keeping
the visualization tractable, although it does not capture the
full structure of the magnetic field. We selected spin periods of 
\Pf, \Po, and $P=0.5$\,ms, with the latter tracing field lines 
that would remain closed for all known pulsars.

Numerically, footpoints lying on the OFLR edge were selected from
field lines that met the integration accuracy criteria as described in 
Section~\ref{subsec:simulation}. For each \texttt{HEALPix} pixel 
belonging to the OFLR, we identified its 
neighbors as pixels sharing at least one vertex 
with the main pixel \citep[see Fig.~3 in][]{Gorski2005}. 
A footpoint was classified as an edge footpoint if at least two of 
its neighbor pixels lay outside the OFLR. This resulted in a 
number of edge lines that was still too large for clean rendering; thus, 
we employed a downsampling procedure. For a coarser \texttt{HEALPix} grid with
$\mathtt{nside}=128$, we took the average of the 3D coordinates of 
footpoints that fell within each coarse pixel and  
calculated the spherical coordinates of the downsampled footpoint by 
normalizing the average vector to unit length. 
%This method provides uniform downsample on a sphere
%regardless of the shape of the OFLR edge curve. 

For a given footpoint grid and field configuration, all field lines were 
integrated to a termination point, either $R_\mathrm{lim}$ or $\RNS$ 
(Section~\ref{subsec:simulation}), but we plotted these lines only 
to the extent of their respective light cylinders. Sometimes more than one OFLR edge
passed precisely through the same pixel regardless of grid resolution, 
indicating a sharp change in the maximum extent of neighboring field 
lines (Fig.~\ref{fig:OFLR_edge_type}). 
%In this case, we added a minuscule jitter to 
%footpoint coordinates and reintegrated the edge lines.

\begin{figure}
\centering
 \includegraphics[width=\columnwidth]{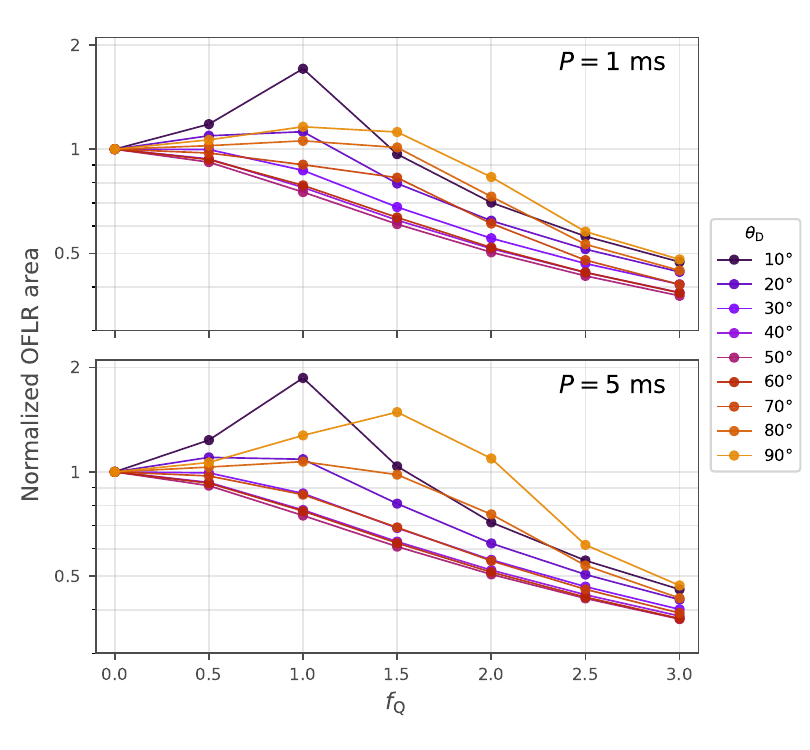}
 \caption{Total surface area of OFLRs as a function of the relative 
 quadrupole strength $\fQ$ for spin periods \Po\ (top) and \Pf\ (bottom). 
 Areas are normalized by the OFLR area of a pure dipole field with 
 the same $\thD$. }
\label{fig:OFL_area}
\end{figure}

\begin{figure*}
\centering
 \includegraphics[width=\textwidth, clip]{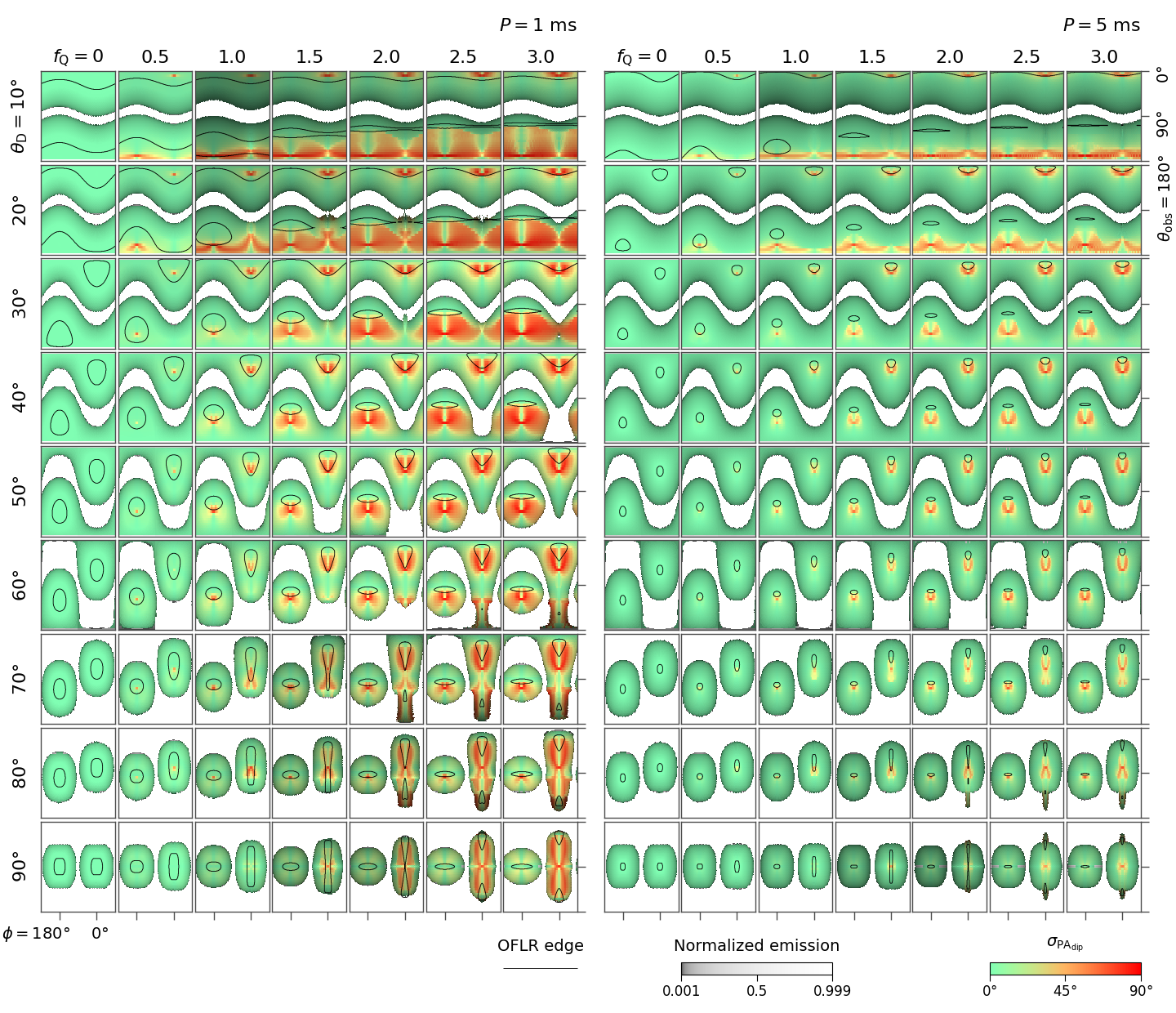}
   \caption{Total radio emission intensity and the RMS PA deviation $\PAdiff$
   from the pure dipole RVM for \Po\ (left) and \Pf\ (right), shown on a grid 
   of $\fQ$ (columns) and $\thD$ (rows). Each panel is a skymap in spin 
   longitude $\phi$ and observer colatitude $\thobs$. Color encodes PA 
   deviation, from green for small values to red for large values, while 
   brightness encodes total emission intensity normalized by the peak of 
   each skymap. The displayed brightness range is strongly compressed 
   between $0.1\%$ and $99.9\%$ of the peak, so small differences in 
   brightness correspond to large intensity variations. Regions below 
   $0.1\%$ of the peak are shown in white. 
   Thin gray contours mark OFLR edges on the star surface.
 }
\label{fig:skymap_PA}
\end{figure*}

\begin{figure*}
\centering
 \includegraphics[width=\textwidth]{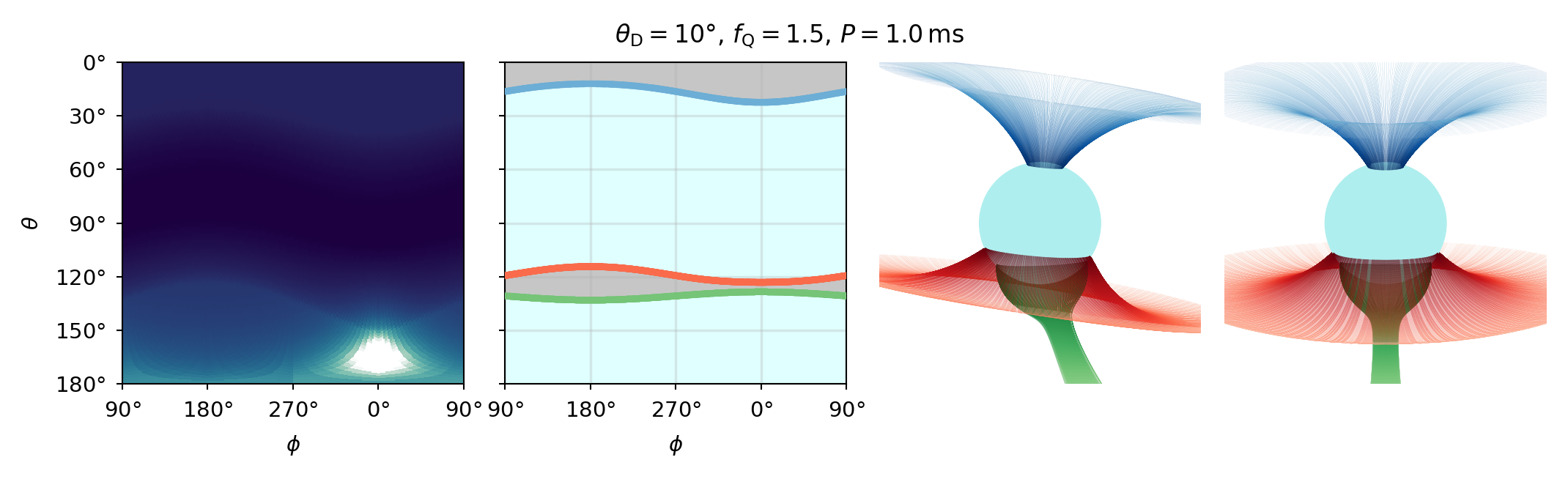}
 \includegraphics[width=\textwidth]{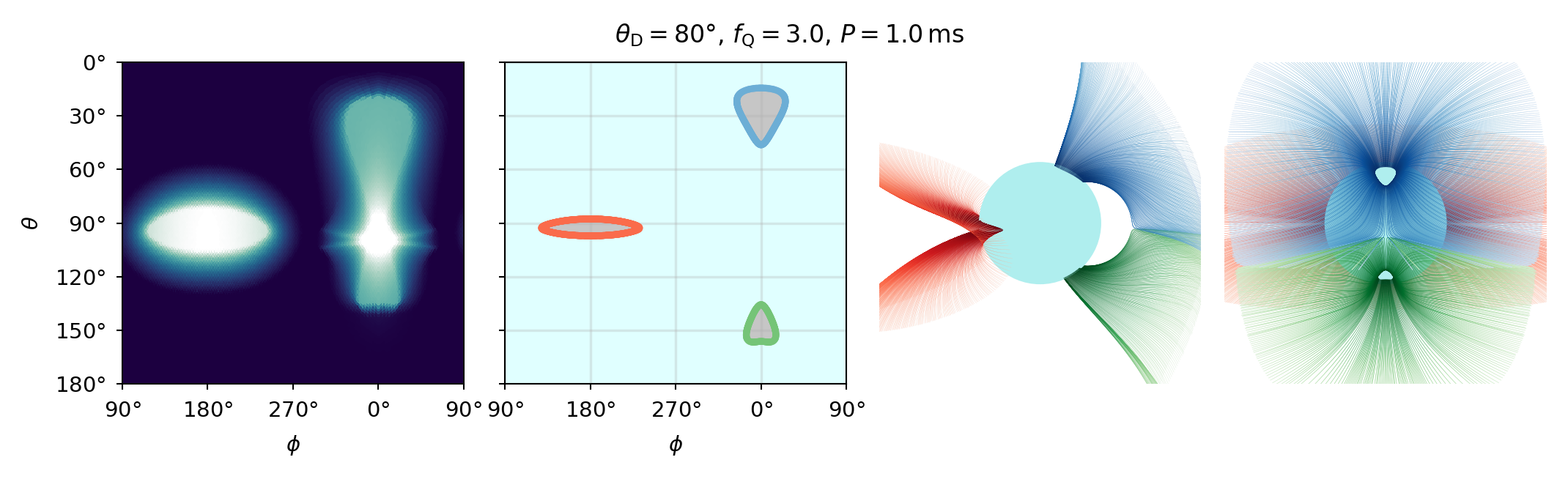}
 \caption{Illustration of the bunching of open field lines that produces localized bright features
in radio emission skymaps for \Po. The \textit{top} row shows a near-aligned configuration ($\thD = 10\degree$,
$\fQ = 1.5$), and the \textit{bottom} row shows a near-orthogonal configuration ($\thD = 80\degree$, $\fQ = 3.0$).
For each configuration, the \textit{leftmost} panel shows the radio emission skymap, while the
\textit{second} panel shows the corresponding OFLRs on the stellar surface.
Emission is produced along field lines whose surface footpoints lie within the OFLRs.
The OFLR edges are color-coded for visualization, and the associated boundary field lines are shown in
the two \textit{rightmost} panels. Further details on each type of plot are provided in the Appendix glossary.
For the near-aligned case, field lines originating in the
 vicinity of the southern boundary of the strip-like OFLR are skewed toward 
 $\phi = 0\degree$, leading to concentrated emission in the skymap. For 
 the near-orthogonal case, field lines emerging from the northern and 
 southern OFLRs approach each other near the rotational equator and spread in 
 azimuth, producing extended bright structures.}
\label{fig:caustics}
\end{figure*}

\begin{figure}
\centering
 \includegraphics[width=0.97\columnwidth, clip]{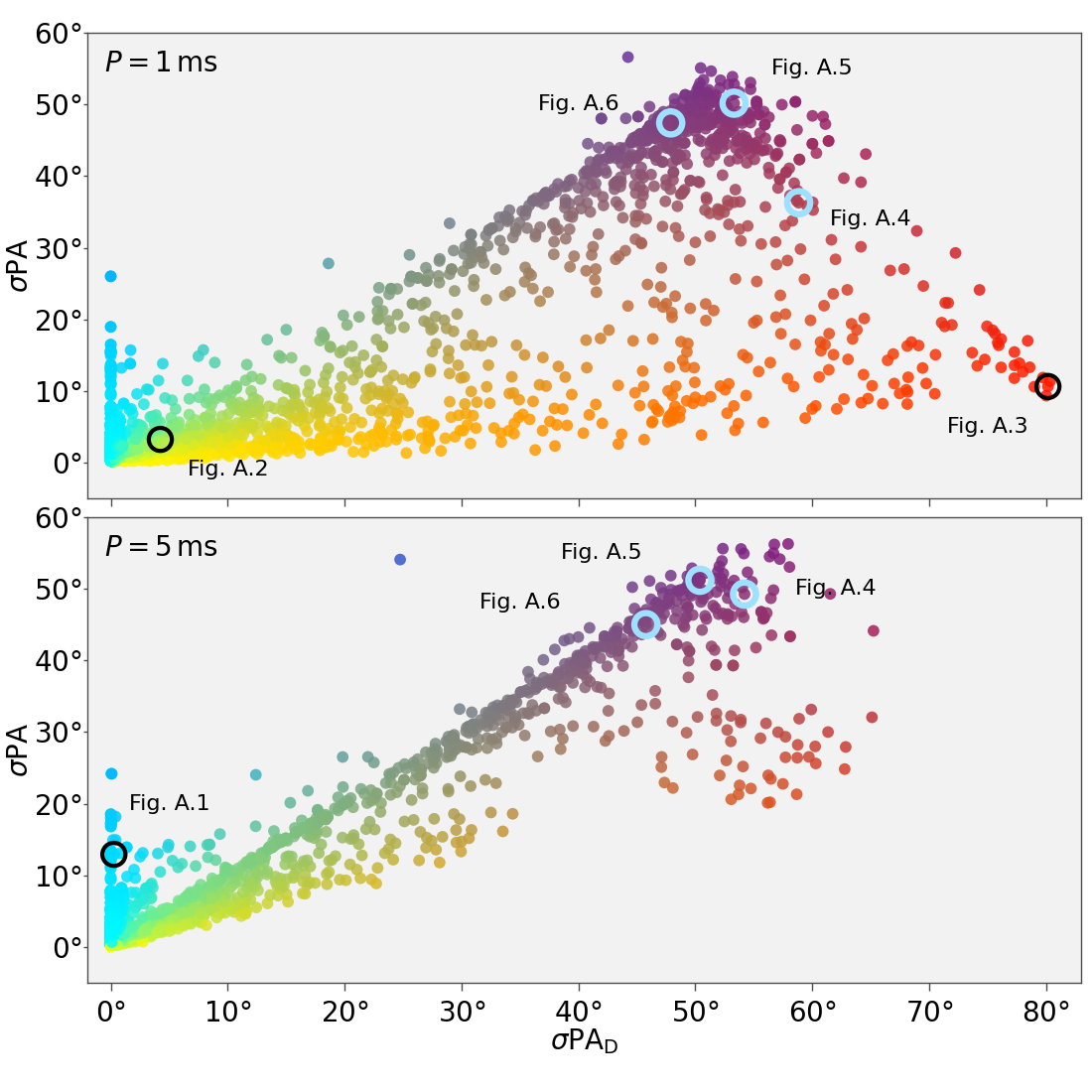}
   \caption{RMS PA deviation with respect to PAs in the pure dipole field ($\PAdiff$, Eq.~\ref{eq:PAdiff})
  and with respect to the actual average PA ($\PAmnrms$, Eq.~\ref{eq:PAmnrms}) 
   for \Po\ (top) and \Pf\ (bottom). Each pair of metric values corresponds to 
   a combination of $\thD$, $\fQ$, and $\thobs$, limited to the $\phi$ bin that 
   corresponds to the maximum $\PAmnrms$. Point color follows the polar 
   RGB mapping in the metrics plane (see text for more details). Open circles 
   mark representative PA examples shown in the Appendix.
 }
\label{fig:pa_metric_scatter}
\end{figure}

\begin{figure*}
\centering
 \includegraphics[width=0.97\textwidth, clip]{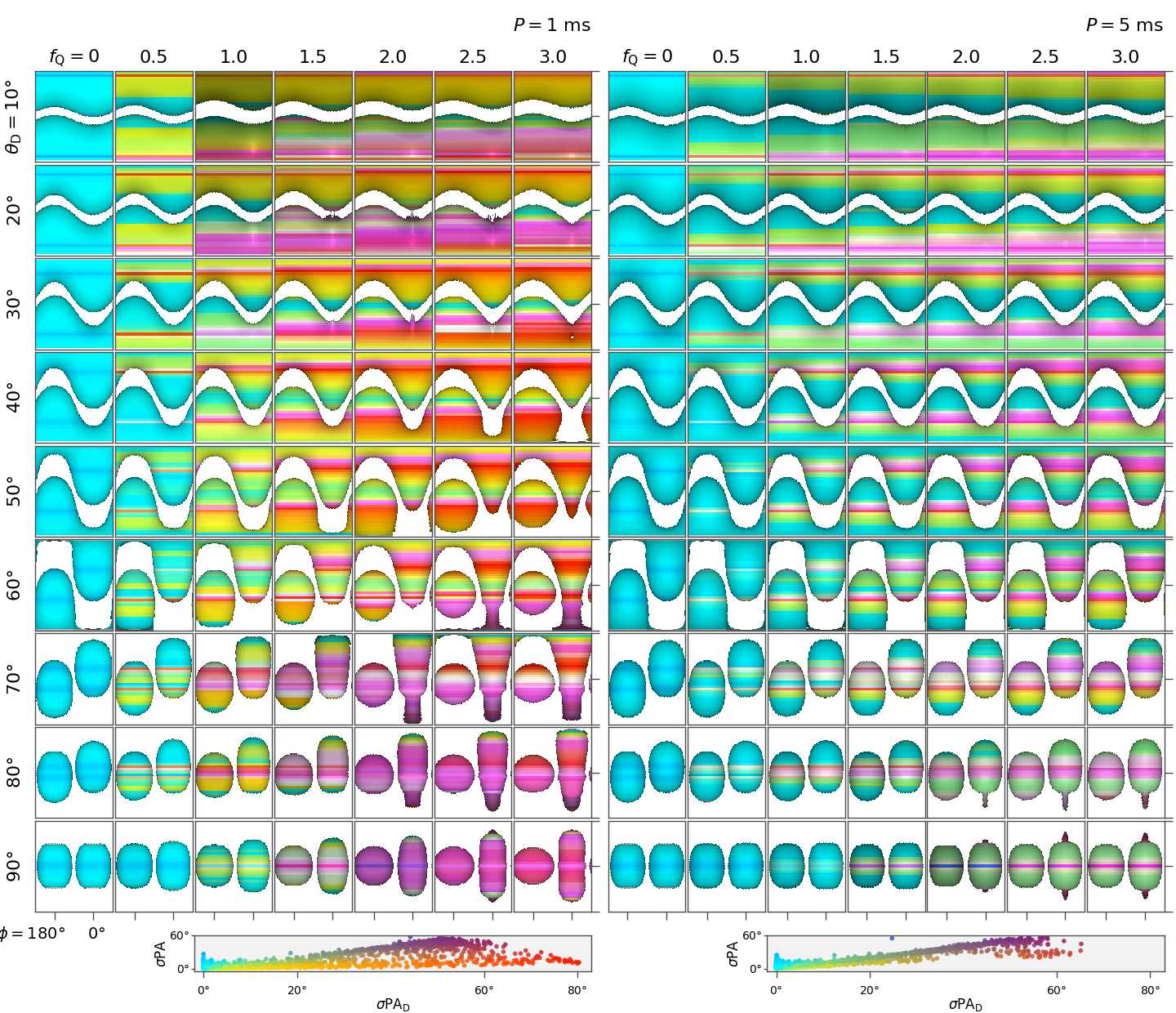}
   \caption{Same skymap grid as Fig.~\ref{fig:skymap_PA}, but with each
$\thobs$ row colored according to the joint values of $\PAdiff$ and $\PAmnrms$.
For each magnetic field configuration and $\thobs$, the color is determined from
the longitude bin with the maximum $\PAmnrms$ and mapped to RGB in the
$(\PAdiff,\PAmnrms)$ plane as described in the text. Brightness encodes the total
emission intensity using the same normalization as in Fig.~\ref{fig:skymap_PA}.
The \textit{bottom} panels show the corresponding $(\PAdiff,\PAmnrms)$ scatter plots for
\Po\ and \Pf.
 }
\label{fig:skymap_metric}
\end{figure*}

With these caveats in mind, we now turn to Fig.~\ref{fig:FL_overview} and describe 
the shape of the OFLRs.
The leftmost column shows the pure oblique dipole field. The OFLRs 
are centered on the magnetic axis and are circular on the stellar 
surface for an aligned rotator, becoming progressively more elongated 
in azimuth as $\thD$ increases.

Adding a spin-aligned quadrupole component 
modifies the OFLR morphology 
in a manner that depends on $\thD$, $\fQ$, and the pulsar period. 
For a near-aligned rotator ($\thD\leq10\degree$) with \Po, the OFLR 
of the pure dipole field encompasses the southern rotational pole. 
As $\fQ$ increases, the northern OFLR shrinks, while the 
southern OFLR expands. At $\fQ=1$, a topological transition occurs 
and an opening develops within the southern OFLR. Inside this 
opening, the quadrupole field locally dominates and the corresponding 
field lines form closed loops close to the stellar 
surface (see also Fig.~\ref{fig:OFLR_edge_type} for an example of such loops). 
The field lines forming the boundary of this opening define a collimated flux 
tube in three dimensions, whose cross-section gradually expands as $\fQ$ increases. 
At the same time, the opening grows in area and the outer boundary of the 
southern OFLR shifts toward the rotational equator. Thus, the southern OFLR 
becomes increasingly narrow in colatitude, adopting a strip-like morphology. 
The northern OFLR does not develop a comparable opening, since the 
dipole and quadrupole field components do not cancel there, and its 
surface area instead decreases monotonically with increasing $\fQ$.

At larger values of $\fQ$, beyond the range explored here, a third 
OFLR appears in the southern hemisphere. With further increase of $\fQ$, 
the location and shape of the three OFLRs gradually approach those 
of a pure, aligned quadrupole field.

For the larger \Pf\ at $\thD=10\degree$, no opening develops
within the southern OFLR. Instead, 
the footprint undergoes a continuous geometric deformation, 
becoming increasingly elongated in longitude and shifting toward
the rotational equator. At large $\fQ$, the resolution of our \texttt{HEALPix} grid becomes 
insufficient to resolve the narrow angular span in colatitude 
of strip-like OFLRs. This produces a string of one-pixel OFLRs.

For intermediate inclination angles, $\thD=30\degree$--$60\degree$ 
on our $\thD$ grid, the qualitative behavior of the OFLRs is 
similar for both pulsar periods considered. In this regime, both 
OFLRs are elongated, but their evolution differs 
between hemispheres. The northern OFLR shifts toward the northern 
rotational pole and becomes increasingly elongated in the 
meridional direction, while the southern OFLR shifts toward 
the rotational equator and elongates primarily in the
azimuthal direction. The northern OFLR gradually develops an 
asymmetric teardrop-like morphology, with its pointed end 
oriented toward the rotational equator.

At $\thD=50\degree$ for \PillA, a third OFLR first appears 
at $\phi=0\degree$. This transition occurs at larger $\thD$ for 
longer periods, appearing for \Po\ at $\thD=60\degree$ and for 
\Pf\ at $\thD=70\degree$.

For \Po\ and larger inclination angles, $\thD=70\degree$--$80\degree$, 
the deformation of the northern OFLR becomes more extreme as $\fQ$ 
increases. The footprint expands strongly in meridional angular span, 
in some cases exceeding $90\degree$, while simultaneously thinning 
at the rotational equator. At 
$\thD=80\degree$, this thinning leads to a topological 
transition in which the northern OFLR splits into two 
disconnected teardrop-shaped regions, each with its pointed 
end oriented toward the rotational equator. As $\fQ$ increases further, 
these regions migrate toward their corresponding rotational poles.

This behavior is most pronounced for the orthogonal case, 
$\thD=90\degree$. Here both OFLRs are symmetric
with respect to the rotational equator. The OFLR centered at $\phi=0\degree$ elongates 
in the meridional direction while thinning at the equator, 
eventually splitting into two teardrop-shaped regions that migrate 
toward the rotational poles as $\fQ$ increases. At the same time, the 
antipodal OFLR at $\phi=180\degree$ elongates predominantly in the 
azimuthal direction and ultimately forms an equatorial strip outside 
the range of $\fQ$ explored here. This qualitative behavior 
is the same for both spin periods.

Thermal emitting regions smaller than the polar cap expected for a pure 
dipole are often interpreted as evidence for multipolar surface magnetic fields.
However, within the parameter range explored here, this was not always
the case. The total OFLR area, normalized by the area of the corresponding
pure dipole OFLR at the same $\thD$, decreased monotonically with increasing
$\fQ$ only for $\thD = 40\degree$--$60\degree$ for \Po\
and for $\thD = 30\degree$--$70\degree$ for \Pf\
(Fig.~\ref{fig:OFL_area}). For $\thD$ closer to $0\degree$
or to $90\degree$, the area remained approximately constant or even increased
for small $\fQ\leq2$.

\section{Emission skymaps and PA behavior}

As the magnetosphere rotates, an observer at colatitude 
$\thobs$ samples emission from different regions of 
the magnetosphere whenever the local field line tangent 
aligns with the line of sight. This produces a spin 
longitude-dependent distribution of radio intensity, 
i.e., a pulse profile, together with an associated 
variation of the PA. 

The goal of this work is not to simulate the shape
of the average profiles, which inevitably depend on
poorly known emissivity profiles along the field lines.
Instead, we explore a geometry-driven umbrella scenario by 
assuming constant emissivity per unit arc length in a broad range of emission
altitudes. The emissivity is uniform across all field lines in the OFLR 
and emission is taken to be completely linearly polarized.

Total intensity skymaps and the associated PA curves were
computed with two different angular resolutions. For 
 emission skymaps, we extract the spherical coordinates of the field line 
tangent vectors at the emission points and map 
them onto the sky using \texttt{HEALPix} binning with 
$\mathtt{nside}=\nsideEmissionMap$, corresponding to the 
angular resolution of approximately $\angResDegEmissionMap\degree$. 
PA curves depend sensitively on the exact coordinates of $\mathbf{t}$,
thus we sampled PAs within a much narrower 
$\delta\theta = \thetaObsTol\degree$ beam, matching  
the assumed angular scale of the elementary emission beam \citep{Gangadhara2004}.
We explored a grid of $\thobs$ spanning 
$[\thobsMin, \thobsMax]$ in steps of $\thobsStep$. 

We start by reviewing the general shapes of emission beams 
together with the bias in PAs that is introduced by quadrupole components.
To quantify the latter, we computed the wrapped angular root-mean-square deviation
between the PA at
each emission point and the PA predicted by the dipolar RVM for the same
$\thD$ (Eq.~\ref{eq:RVM}):
\begin{equation}
\begin{aligned}
& \PAdiff = \left[\frac{1}{N}\sum_{i=1}^{N}\left(\mathrm{wrap}(\Delta\mathrm{PA}_i)\right)^2\right]^{0.5},\\
& \Delta\mathrm{PA}_i = \mathrm{PA}_i- \arctan\left(\dfrac{\sin\thD\,\sin\phi}
 {\sin\theta_t\cos\thD - \cos\theta_t\sin\thD\cos\phi}\right).
\end{aligned}
\label{eq:PAdiff}
\end{equation}
Here $N$ is the number of emitting points within each $\paPhiBinWidth$ bin
in spin longitude $\phi$ and the wrap of an angle $a$ is defined as follows:
\begin{equation}
\mathrm{wrap}(a) = [(a + 90\degree) \, \mathrm{mod}\, 180\degree ] - 90\degree.
\end{equation}

In Eq.~\ref{eq:PAdiff} we used the actual colatitude of the field
line tangent $\theta_t$ (Eq.~\ref{eq:emission_point}) instead of $\thobs$
for the following reason:
for a pure dipole, when $\thobs = \thD$, 
$\mathrm{PA}_i$ for points close in $\phi$ to $0\degree$ or $180\degree$
depends sensitively on the exact value of $\theta_t$, resulting in
a large spread of points around the $180\degree$ jump in the analytical RVM curve
(see an example in Fig.~\ref{fig:PA_scenario_D}). Consequently,
using $\thobs$ instead of $\theta_t$ produces a large
$\PAdiff$ in the corresponding $\phi$ bins even for $\fQ=0$.

Figure~\ref{fig:skymap_PA} presents an overview of the resulting emission
skymaps and $\PAdiff$ patterns for two representative spin periods,
shown as a function of observer colatitude $\thobs$ and magnetic field
configuration. In each panel, color corresponds to the magnitude of
$\PAdiff$, while slight variation 
in brightness encodes the total emission intensity. 
This combined representation allows
changes in PA behavior to be related directly to the morphology and relative
intensity of the corresponding emission beam.

The overall morphology of the emission beams in Fig.~\ref{fig:skymap_PA} varies 
systematically with magnetic field configuration. For $\fQ=0$, the skymaps consist
of two opposing beams centered on the magnetic axis. The shape of each beam
on the sky follows the shape of the corresponding
OFLR footprint on the stellar surface: it is circular for $\thD=0\degree$
and becomes progressively compressed in colatitude relative to longitude
as $\thD$ increases. Introducing a quadrupole component breaks this direct
correspondence. For $\fQ>0$, the morphology of the emission beam is no longer
a simple reflection of the OFLR footprint, although for emission restricted
to a narrow range of heights close to the stellar surface it follows more closely
the OFLR morphology. For \Pf, the beams integrated
over the full adopted range of emission heights remain broadly dipole-like
throughout the explored range of $\fQ$.

The grid of magnetic field configurations explored in this work revealed 
two regimes of substantial emission collimation, in which multiple field 
lines bunch together and produce distinct bright spots in the skymaps. 
Because each skymap is normalized by its brightest pixel, the presence of 
such a localized intensity maximum suppresses the relative brightness 
of the remaining emission. Skymaps containing strongly collimated spots 
can therefore be readily identified by their overall darker appearance.

Both bunching regimes are illustrated in Fig.~\ref{fig:caustics}.
The first regime occurs at relatively small dipole inclination angles,
for $\thD < 40\degree$ for \Po\ and $\thD<20\degree$ for \Pf.
It is manifested as a pronounced collimation of emission in the southern hemisphere,
opposite in spin phase to the dipolar beam. This arises from the azimuthal skewing of
field lines originating in the southern OFLR, which concentrates emission from a
broad range of field lines into a narrow range of spin longitudes.

The second regime occurs at larger inclination angles,
$\thD>50\degree$ for \Po\ and $\thD>70\degree$ for \Pf.
Emission associated with field lines originating from the northern OFLR becomes
increasingly elongated in longitude. For fixed $\thD$, increasing $\fQ$ causes
the OFLR to narrow near the rotational equator and eventually split into two
teardrop-shaped regions. Field lines originating near the narrow equatorial section,
or near the equator-facing tips of the separated regions, fan out coherently toward
the rotational equator. Their tangent directions therefore span an extended range
of spin longitude while remaining confined to a relatively narrow range of
colatitude, producing elongated stripes in the skymaps.

We now turn our attention to $\PAdiff$. The first noticeable feature is that
regions of large PA deviation have a characteristic lung-like shape, with
smaller $\PAdiff$ near $\phi=0\degree$ and $\phi=180\degree$. This arises
because both terms entering $\Delta\mathrm{PA}_i$ in Eq.~\ref{eq:PAdiff} are generally smaller in
these regions, and consequently their difference is also reduced.
This behavior may not persist for more general magnetic configurations
in which multipole moments and spin vector do not all
lie in the same plane. However, such configurations fall outside the parameter
space explored here.

Increasing spin period shrinks regions of large $\PAdiff$ both in $\phi$
and in $\thobs$. For both spin periods explored in this work, larger emission
altitudes are associated with smaller $\PAdiff$. In most cases, $\PAdiff$ is
also reduced near the edges of the emission beams.

The $\PAdiff$ provides a compact measure of how
strongly the PA behavior departs from the dipolar RVM, but it does not
capture the detailed structure of the PA distribution. A large $\PAdiff$
may correspond, for example, to a single compact PA track that differs from the
dipolar RVM prediction, to a broad scatter of PA points, or to anything
in between. To quantify the spread of PA values within each bin, we
computed the wrapped RMS deviation from the circular mean PA in the same bin:
\begin{equation}
\begin{aligned}
& \PAmnrms = \left[\frac{1}{N}\sum_{i=1}^{N}(\mathrm{wrap}\left(\mathrm{PA}_i - \mu_\mathrm{PA}\right))^2\right]^{0.5},\\
& \mu_\mathrm{PA} = \dfrac{1}{2}\mathrm{atan2}\left(
\dfrac1N\sum_{i=1}^{N}\sin2\mathrm{PA}_i,\ 
\dfrac1N\sum_{i=1}^{N}\cos2\mathrm{PA}_i
\right),
\end{aligned}
\label{eq:PAmnrms}
\end{equation}
where $\mathrm{atan2}$ is the \texttt{numpy} arctangent function that selects the
correct quadrant based on the signs of its two arguments.

Both $\PAmnrms$ and $\PAdiff$ depend on $\phi$ and can vary substantially across 
the pulse phase. However, for a given field configuration and $\thobs$, we are 
interested in the behavior of the PA track as a whole. For the subsequent analysis, 
we therefore use the values of $\PAmnrms$ and $\PAdiff$ from the phase bin in which 
$\PAmnrms$ is maximal. This characterizes the largest PA spread across the on-pulse 
window and the corresponding deviation of the bulk of the PA points from the purely dipolar case.

Figure~\ref{fig:pa_metric_scatter} shows the scatter plot of $\PAmnrms$ and $\PAdiff$ for
all field configurations and $\thobs$. The points form several broad groups, but these
groups are not sufficiently distinct to motivate formal clustering. Instead, we used a
continuous color mapping based directly on the two metrics. Each point was assigned an
RGB tuple in which the green channel decreases linearly with radius
$\rho=(\PAdiff^{2}+\PAmnrms^{2})^{0.5}$,
from unity at the origin to zero at $\rho=\paMetricRhoGZero$.
The balance between the red and blue channels depends on the polar angle
$\tan\varphi=\PAmnrms/\PAdiff$:
the red channel remains at its maximum possible value for $\varphi\leq \paMetricPhiRMax$, 
while the blue channel is zero; the blue channel remains at its maximum possible value 
for $\varphi\geq \paMetricPhiBMax$, while the red channel is zero; between these limits, 
the two channels are interpolated linearly with respect to $\varphi$.

This mapping gives the colors a direct relation to the two PA metrics. At large
$\rho$, red points correspond to relatively compact PA tracks that differ strongly
from the dipolar RVM, while blue points correspond to cases in which $\PAmnrms$ is
large relative to $\PAdiff$. Purple points occupy the intermediate region, where both
metrics are substantial, indicating several PA tracks or large scatter of PA values. 

As $\rho$ decreases, the increasing green channel contribution makes
the colors progressively brighter. At small values of both metrics, points range
from yellow when $\PAdiff$ dominates, through green for the intermediate ratios, to cyan
when $\PAmnrms$ dominates. 

We discuss these regimes in more detail below, using the same color mapping overplotted
on skymaps with the same brightness distribution as in Fig.~\ref{fig:skymap_PA}.
These plots, shown in Fig.~\ref{fig:skymap_metric}, provide an overview of
the PA behavior associated with each magnetic field configuration and LOS colatitude.

For $\fQ=0$, $\PAdiff$ is very small in most cases, with median values of $0.01\degree$
for \Po\ and $0.001\degree$ for \Pf. In these pure dipole cases, $\PAmnrms$ is generally
larger than $\PAdiff$, producing a vertical cyan column along the leftmost edge of the scatter
point cloud. This difference follows from the definitions of the two metrics.
$\PAmnrms$ is computed with respect to a single circular mean PA within each longitude
bin, whereas $\PAdiff$ compares each individual PA point with the dipolar RVM prediction
evaluated using the corresponding tangent direction. Finite PA variation within a
longitude bin can, therefore, increase $\PAmnrms$ while leaving $\PAdiff$ very small.

In some extreme pure dipole cases, $\PAmnrms$ reaches values of order $10\degree$,
shifting these points toward darker blue. The corresponding regions appear as blue
stripes in the skymaps for $\fQ=0$. This occurs in two situations. In the first,
$\thD = \thobs$, and the PA points are scattered across the $180\degree$ jump.
Because $\PAdiff$ is computed using $\theta_t$, this scatter does not increase
$\PAdiff$, but it does increase $\PAmnrms$ (Fig.~\ref{fig:PA_scenario_D}).
The second situation occurs when the PA gradient is large within a
$\paPhiBinWidth$ bin. This contribution to $\PAmnrms$ could be reduced by replacing
$\mu_\mathrm{PA}$ with a linear fit to the PA values within the bin. However, such a
fit becomes unstable for the more complex PA behavior discussed below, so we retain
the circular mean.

For $\fQ>0$, cyan regions occur at small $\fQ=0.5$ for \Po\ and throughout the explored
range of $\fQ$ for \Pf. For both periods, they tend to occur closer to the edges of
the emission beams.

A particularly interesting part of the distribution is the bright yellow to green region, where both
metrics are smaller than $\sim 5\degree$. In this regime, PA tracks remain essentially
independent of emission altitude and closely resemble those of a pure dipole despite the
presence of a quadrupole component (see Fig.~\ref{fig:PA_scenario_A} for a representative
example). Such regions occur more frequently for \Po\ than for \Pf.

As both metrics increase, the contribution of the green channel gradually decreases, and
the points trace a clear progression from red through purple toward blue along the outer
edge of the distribution. However, there are no cases with large $\PAmnrms$ and small
$\PAdiff$, which would occupy the blue end of this sequence. Such cases could arise 
for very large PA gradients within a single longitude bin. The opposite behavior
does occur for \Po: some points have large $\PAdiff$ but relatively small $\PAmnrms$, and are
therefore dominated by the red channel. In these cases, the PA track remains relatively
tight but does not follow the dipolar RVM. It is never as tight as in the pure dipole case
(see Fig.~\ref{fig:PA_scenario_E} for an example) and still shows some dependence on
emission altitude.

For \Po, such PA behavior is typical at small $\thD$ and for
$\thobs\approx\thD$, with $\thD < 50\degree$. At larger $\thD$, $\PAmnrms$ can remain
small even when additional PA tracks appear. These additional tracks contain relatively few
emission points and therefore contribute little to $\PAmnrms$. For \Pf, this type
of behavior is generally absent.

Cases in which both metrics are relatively large, corresponding to the purple region,
comprise two distinct types of behavior. In the first, $\PAdiff$ is large and the PA values
depend on emission altitude. Within narrow altitude ranges, the PAs form S-shaped curves
resembling pure dipole RVM tracks, with their shape evolving gradually with emission altitude
(Fig.~\ref{fig:PA_scenario_B}). If the observed emission originates from a localized region
of the magnetosphere, applying a dipolar RVM fit may, therefore, lead to an incorrect inference
of $\thD$.

In the second case, the PA behavior depends strongly on which region of the OFLR contributes
to the emission (Figs.~\ref{fig:PA_scenario_C_Psmall}--\ref{fig:PA_scenario_C_Pbig}).
These cases are associated with strong field line bunching, which brings emission from
distinct regions of the OFLR into overlapping ranges of spin longitude. A single dipolar
PA track may then fail to provide an adequate description of the observed PA structure,
depending on the extent of the active field lines and the range of emission altitudes.

\section{Biases in dipolar RVM fits to quadrupole-modified PA curves }

\begin{figure*}
\centering
\includegraphics[width=0.46\textwidth, clip]{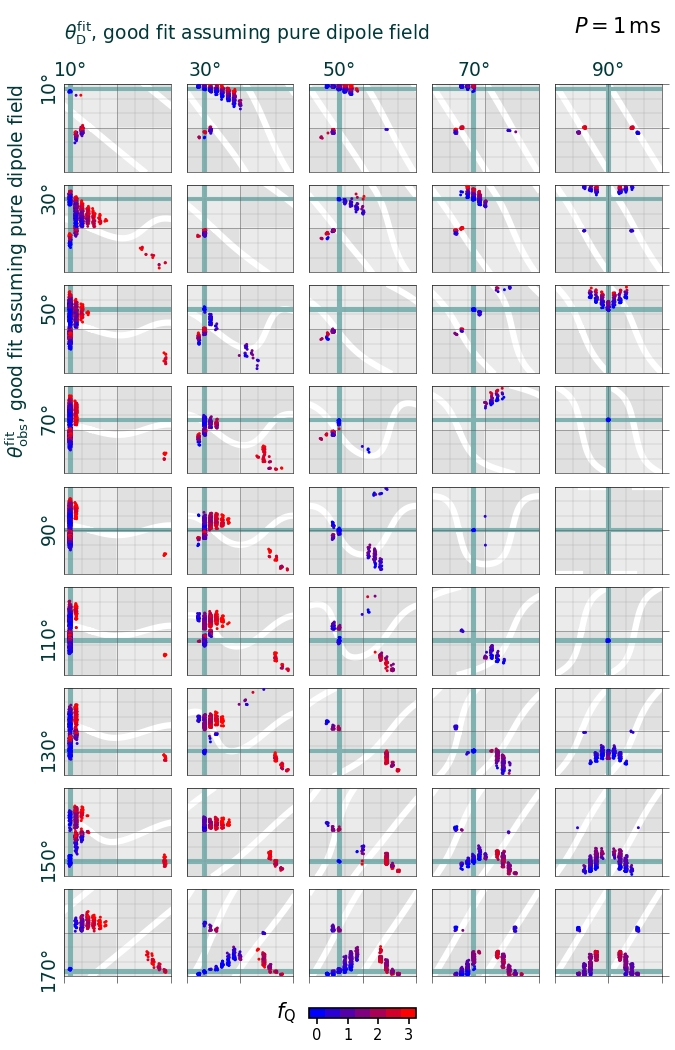}\includegraphics[width=0.46\textwidth, clip]{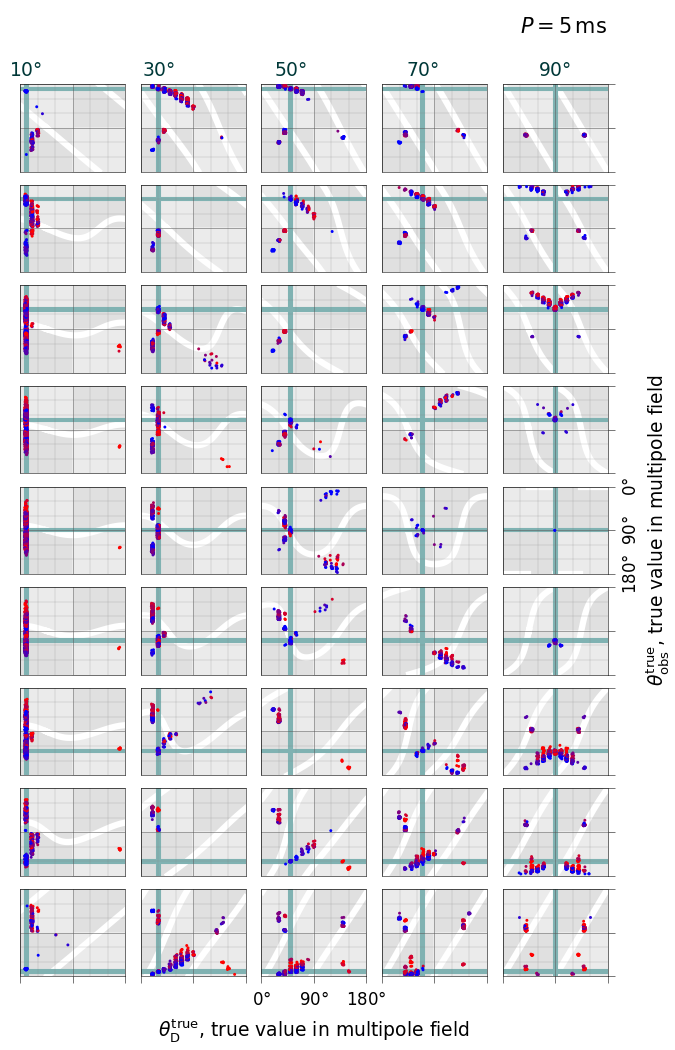}
\caption{Distribution of true magnetospheric geometries associated with selected 
dipolar RVM fits for \Po\ (left) and \Pf\ (right). 
Each panel is defined by the fitted angles $(\thD^\mathrm{fit},\thobs^\mathrm{fit})$, 
given by its column and row, and shows the corresponding  
$(\thD^\mathrm{true},\thobs^\mathrm{true})$ pairs as points, colored 
by $\fQ^\mathrm{true}$. The green vertical and horizontal bands, each $5\degree$ wide, 
mark $\thD^\mathrm{fit}$ and $\thobs^\mathrm{fit}$, respectively. True angle pairs 
are jittered by the same amount along both axes for visibility. 
The white curve overlaid in the background shows the PA curve predicted by 
the corresponding fitted RVM as a visual reference. Its placement within the panel 
is schematic and does not correspond to the $(\thD^\mathrm{true},\thobs^\mathrm{true})$ coordinate system.
}
\label{fig:blind-fit-thd-scatter-grid}
\end{figure*}

A number of recent works aiming to constrain magnetospheric
geometry assume that the magnetic field at the emission altitude is essentially
dipolar and fit the observed PA curves accordingly \citep{Petri2025,Petri2026,Qi2026}.
It is therefore useful to ask whether PA samples produced by a multipolar field
can still admit acceptable dipolar RVM fits while yielding biased estimates 
of $\thD$ and $\thobs$.

The analysis in the previous section established that in the range of
magnetic field configurations explored here the emission accumulated in a broad range of heights
does not exhibit a tight PA curve different from the one produced 
by the dipole component only. However, in more realistic models of pulsar radio emission,
the emission region is expected to be more localized, both to a subset of field
lines and, within curvature radiation models of pulsar radio emission, to a
limited range of altitudes \citep{Gil2000,Gangadhara2004}. The difficulty is that neither the active field lines
nor the relevant emission altitudes are known a priori, particularly for a
multipolar field with a more complex topology than an oblique dipole.
We therefore make a substantial simplification by dividing the previously adopted
emission region into altitude slices, and 
within each slice treating all field lines in the OFLR as equally emitting.

More specifically, we divided the adopted emission altitude
range $[\rMinEmiss\RNS,\ \rTrimCoeff\RLC]$ into
\blindFitNHeightChunks\ equal altitude slices and treated each slice separately.
For each slice, we fitted a pure dipolar RVM (Eq.~\ref{eq:RVM})
over a trial grid in $(\thD,\thobs)$,
using the wrapped RMS difference between the numerically
integrated PAs and those predicted by Eq.~\ref{eq:RVM} as the fit statistic:
\begin{equation}
\begin{aligned}
& \PAslice = \left[\frac{1}{N}\sum_{i=1}^{N}\left(\mathrm{wrap}(\Delta\mathrm{PA}_i)\right)^2\right]^{0.5},\\
& \Delta\mathrm{PA}_i = \mathrm{PA}_i- \arctan\left(\dfrac{\sin\thD\,\sin\phi}
 {\sin\thobs\cos\thD - \cos\thobs\sin\thD\cos\phi}\right).
\end{aligned}
\label{eq:PAslice}
\end{equation}

Our set of numerically integrated magnetospheric configurations did not extend 
beyond $\thD = 90\degree$. For the centered oblique dipole plus spin-aligned 
quadrupole considered here, configurations with $\thD > 90\degree$ are equivalent 
to those with $180\degree-\thD$ after a $180\degree$ shift in azimuth. 
The corresponding magnetic field lines therefore have the same geometry, with 
their locations shifted by $180\degree$ in spin phase. The same symmetry 
applies to the pure dipolar RVM, for which the PA curve at $\thD > 90\degree$ 
is identical to that at $180\degree-\thD$, shifted by $180\degree$ in spin 
longitude. We therefore extended the fitted $\thD$ grid to $(90\degree,\ 170\degree]$.

\begin{figure}
\centering
\includegraphics[width=\columnwidth, clip]{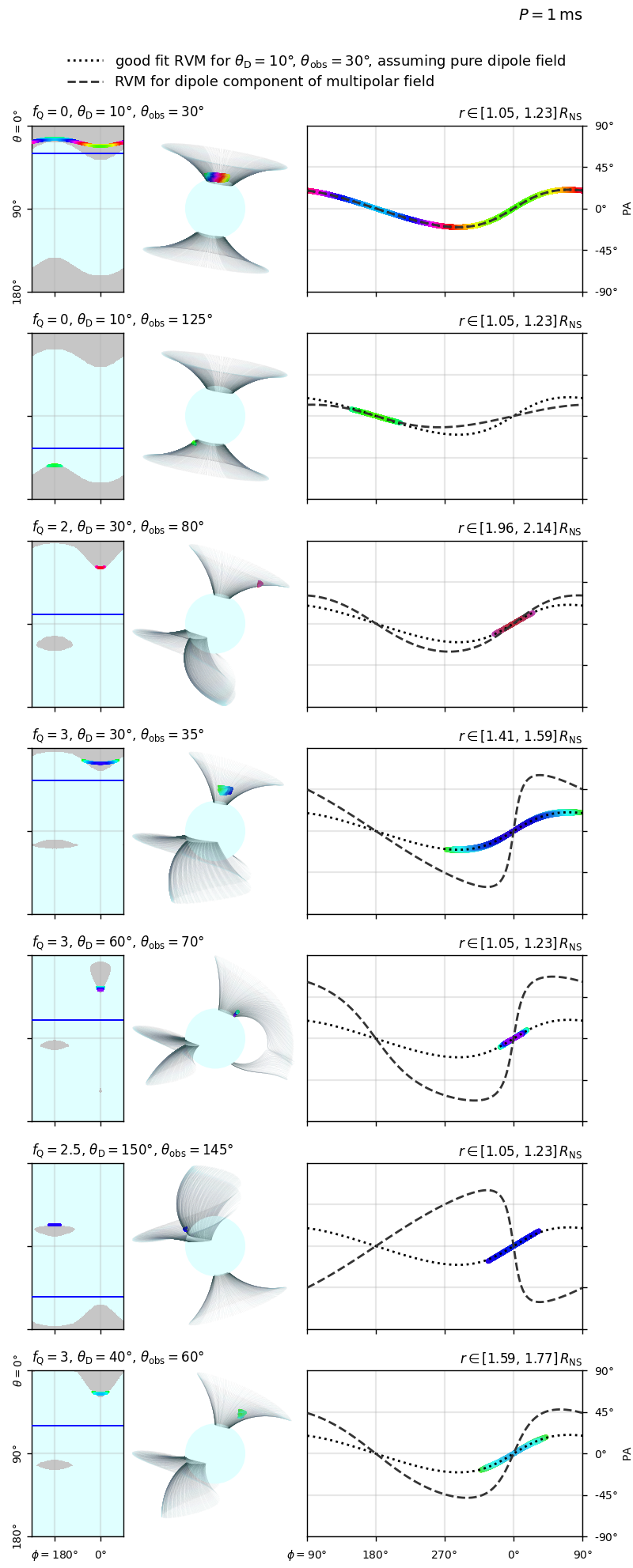}
\caption{Examples of PA curves well fit by a pure dipolar RVM with
$\thD=10\degree$, $\thobs=30\degree$ at \Po.
Emission is restricted to one of \blindFitNHeightChunks\ equal altitude slices spanning
$\rMinEmiss\RNS$ to $\rTrimCoeff\RLC$.
The top two rows show pure dipole configurations, while the lower rows include a quadrupole component.
Panel types and plotting conventions follow the Appendix glossary. In the PA panels, the dotted curve shows 
a good fit for pure dipolar RVM with $\thD=10\degree$, $\thobs=30\degree$, while the dash-dotted 
curve shows the RVM corresponding to the dipolar component of the underlying multipolar field.}
\label{fig:blind-fit-examples}
\end{figure}

As our goal was to identify examples that closely resemble RVM curves, rather than to obtain
a single set of best-fit parameters,
for subsequent analysis we selected all configurations that produced a small
$\PAslice<\blindFitRmsPlotMax\degree$. We further required that emission was present
over an on-pulse window of at least $\blindFitPhiBinFracMin\times360\degree$,
where the on-pulse window was defined as the union of $1\degree$ spin longitude bins occupied
by the selected points. This requirement was imposed because PA curves confined to
very narrow windows result in a large spread of acceptable $(\thD,\thobs)$ pairs.

The fits resulted in a collection of true and fitted magnetospheric geometry pairs,
\begin{equation}
\begin{aligned}
&\mathrm{gm^{true}} = (\fQ^\mathrm{true},\ \thD^\mathrm{true},\ \thobs^\mathrm{true}),\\
&\mathrm{gm^{fit}} = (0,\ \thD^\mathrm{fit},\ \thobs^\mathrm{fit}).
\end{aligned}
\end{equation}

We can now examine the distribution of possible $\mathrm{gm^{true}}$ values associated with 
a given fitted geometry $\mathrm{gm^{fit}}$. Such distributions are shown in 
Fig.~\ref{fig:blind-fit-thd-scatter-grid}
for a set of representative $\mathrm{gm^{fit}}$ pairs, with $\fQ^\mathrm{true}$ encoded
by color. Each panel shows, for a given pair of fitted angles, the distribution of true 
geometries that can produce the RVM curve plotted in the background. 
It should be kept in mind, however, that because we restrict the emission to a narrow range of
heights, the resulting PA curves usually do not cover the full extent of the RVM curve in $\phi$,
but are instead concentrated around $\phi=0\degree$ and $\phi=180\degree$.

Also, a few remarks should be made. First, not all $(\thD^\mathrm{true},\ \thobs^\mathrm{true})$
are recovered by our fit even when $\fQ^\mathrm{true}=0$.
There are two reasons for this. When $\thD=\thobs$, $\PAslice$ exceeds our selection threshold because
small discrepancies between $t_\theta$ and $\thobs$ translate
into large variations in PA (Fig.~\ref{fig:PA_scenario_D}).
In addition, when $\thobs$ is far from $\thD$, the LOS may not
cross the emission beam, or the on-pulse window may fail to satisfy our extent threshold.

Second, even at $\fQ^\mathrm{true} = 0$, the distribution of $\PAslice$ statistics
does not exhibit a sharp minimum, but instead shows similar values over broad ranges of
$(\thD^\mathrm{fit},\ \thobs^\mathrm{fit})$. This occurs because, for a narrow range 
of emission altitudes, PA is restricted to relatively narrow
on-pulse windows. Within these windows, PA curves from different geometries may overlap within
our $\PAslice$ tolerance. This is a known degeneracy of RVM fitting \citep{Johnston2023}.

The presence of a quadrupole component generally broadens the
range of true angle pairs, meaning that multipolar configurations spanning
a wider range of geometries can be mistaken for a specific dipolar
configuration. The extent and shape of this broadening depend strongly on
the observed RVM curve and on the spin period.

To illustrate how PA curves from different multipolar configurations can all
be well described by the same fitted dipolar geometry, we selected
$(\thD^\mathrm{fit},\thobs^\mathrm{fit})=(10\degree,30\degree)$ and show emission
points and PA curves for several representative $\mathrm{gm^{true}}$ configurations in
Fig.~\ref{fig:blind-fit-examples}. The corresponding footpoints of the active field
lines, last open field lines, emitting points, and PA curves are shown in the same figure.
The top row shows the pure dipole case with
$(\thD,\thobs)=(10\degree,30\degree)$. The second row also shows a pure dipole
configuration, but with a different $\mathrm{gm^{true}}$ that nevertheless produces a
similar PA curve within the on-pulse region. The remaining rows show multipolar
configurations whose PA curves are also well described by the RVM with
$(\thD,\thobs)=(10\degree,30\degree)$.

\section{Discussion}

In this work, we explored how the presence of a global quadrupole component modifies PA
behavior within the RVM framework for pulsars with small
light cylinders. Even within the restricted case of an aligned axisymmetric quadrupole,
we find a wide range of PA morphologies, spanning nearly dipole-like behavior
to strongly altitude- and region-dependent PA structures. These results demonstrate
that the relation between the large-scale magnetic field geometry and the observed
PA curve is not unique, even within the relatively simple magnetic configuration
considered here.

At present, a direct comparison with observed MSP PA curves is premature.
The results presented here represent an umbrella of geometrical possibilities
rather than a predictive model for a specific source. In real pulsars, only a
subset of the open field lines is likely to be active. Introducing a quadrupole
component substantially alters not only the topology of the open field line
regions, but also the surface magnetic field strength, field orientation
relative to gravity, and local field line curvature. All of these factors
may influence particle acceleration and pair production, and therefore the
location and extent of the radio emission region. In particular, the more
complex curvature structure of quadrupole-modified field lines should be
taken into account when constraining plausible emission altitudes.

Observed MSP PA curves introduce an additional complication. They frequently
contain sharp changes, small-scale irregularities, and structures that do not
resemble smooth RVM sweeps \citep{Curylo2026}. Such structures cannot be reproduced 
by the large-scale magnetic geometry considered here. Propagation through
the magnetospheric plasma is therefore likely to be important \citep{Sur2024}. 
Moreover, an average PA curve does not necessarily
represent a single underlying PA track. Individual pulses may populate a
distribution of PA values, while averaging preferentially weights the brightest
parts of this distribution. Single-pulse polarization observations would therefore
provide a more direct test of the multiple PA tracks found in our calculations.
Such studies remain scarce for MSPs, largely because of their low single-pulse
signal-to-noise ratios \citep{Palliyaguru2023}.

Despite these limitations, several qualitative conclusions can be drawn.
First, PA curves resembling the classic S-shaped RVM swing are common even
in the presence of a substantial quadrupole component. Consequently, for a
fortuitous position of the observer's LOS, the observation of a dipole-like
PA curve does not by itself rule out significant multipolar structure in the
magnetosphere. This provides a possible configuration in which the radio PA
is well described by a dipolar RVM while the surface field geometry, and
therefore the thermal X-ray emitting regions, differ substantially from
that expected for a pure dipole.

Second, multipolar fields naturally produce multiple PA tracks associated
with emission from different regions of the OFLR and from different emission
altitudes. If only a restricted range of field lines and altitudes is active,
an individual observed track can still resemble an RVM curve, but a fit to it
may recover an incorrect value of $\thD$. If different radio frequencies sample
different subsets of field lines or emission altitudes, the same effect can
produce frequency-dependent changes in PA morphology. 

Our calculations cover very small values of $P$ and a limited range of quadrupole
strengths. The comparison between \Po\ and \Pf\ suggests that the
effects of the quadrupole component on the radio emission geometry generally
become less pronounced as $P$ increases. With increasing $\RLC$, the OFLR
occupies a smaller fraction of the stellar surface, while the quadrupole
component decreases more rapidly with radial distance than the dipole component.
Nevertheless, the location and morphology of the OFLR can remain substantially
different from those of a pure dipole. PA deviations may therefore persist if
radio emission originates sufficiently close to the star.

Our results also have implications for the interpretation of thermal X-ray
emission. Emitting areas inferred from blackbody luminosity fits are often
comparable to or smaller than the open field line region expected for a pure
dipole field \citep{Bogdanov2013,Bogdanov2006,Webb2019}. Smaller emitting areas
are commonly attributed either to non-uniform heating within the open field line
region or to the presence of an offset dipole or higher-order magnetic multipoles
\citep{Zhang2003a,Harding2022}. Our calculations show that the presence of a
global quadrupole component does not in general imply a reduction of the OFLR
area. In some regions of parameter space, the total OFLR area instead exceeds
that of the corresponding pure dipole. This may also be relevant to sources for
which inferred thermal emitting regions are larger than their nominal dipolar
OFLRs \citep{Miller2026,Dittmann2024}.

A natural extension of this work would be to replace the broad emission altitude
range used here with a prescription for the active radio emission region. Such a
prescription could be tied to local field line curvature and pair production, or
introduced phenomenologically. Synthetic pulse profiles could then be used to
study the resulting distribution of profile widths and its dependence on spin
period and magnetic geometry, and to compare these results with predictions for
a pure dipole field \citep{Johnston2019}. As a complementary approach, Stokes
tomography could be used to confront the model with observations by considering
the full structure in $Q$ and $U$ rather than PA alone
\citep{Chung2011a,Chung2011b}.

A more ambitious extension would require a self-consistent magnetospheric model
in which the magnetic field, currents, surface heating, and radio polarization
are treated jointly. One possible approach would combine a plasma-filled
magnetosphere with physically calculated return currents and thermal X-ray
surface emission \citep{Huang2025}, constrained directly by NICER observations,
with radio emission generated in the same magnetosphere and propagated through
the plasma \citep{Sur2024}, with inclusion of all relativistic effects. 
Such a calculation would provide a direct test
of whether a single magnetic configuration can simultaneously reproduce the
thermal X-ray morphology and radio polarization behavior.

Finally, the existence of multiple PA tracks from different emitting field lines
suggests possible tests outside the MSP population. Single-pulse PA distributions
in normal pulsars may retain information about which parts of a multipolar
magnetosphere are active, particularly when highly linearly polarized pulses can
be isolated. Similar considerations may also be relevant to magnetars and Fast Radio Bursts (FRBs).
\citet{Liu2026} reported approximately Gaussian distributions of PA around
repeater-dependent mean values for FRBs. If at least some FRB emission originates
on closed magnetic field lines in magnetars, multipolar magnetic structure could
provide an additional source of location-dependent PA variation.

\section*{Acknowledgments}
Some of the results in this paper have been derived using the \texttt{healpy} and \texttt{HEALPix} packages.
AVB thanks Vladislav Kondratiev for discussions and comments on the manuscript, 
and Weiyang Wang for proofreading the manuscript.
The initial version of \texttt{psr-magpy}, all analytical derivations, the literature study, 
and the scientific content of the manuscript were prepared by AVB. 
Artificial intelligence tools, including Cursor AI with various underlying models (years 2024$-$2026), 
were used to improve the field line integration code and to prepare code for the figures. 
AI-assisted modifications to the numerical code were checked against analytical 
limiting cases and by direct inspection of representative numerical results. 
Code used for data visualization was verified against the underlying numerical 
arrays and representative limiting cases. ChatGPT (various models, same years) 
was used for language editing, as well as for natural language literature searches 
and for identifying possible logical inconsistencies in the final version of the 
manuscript.

\begin{appendix}
\section{Figures and panel glossary}
\label{app:figures}The figures in this Appendix provide detailed views of representative
PA behaviors discussed in the main text and use the same diagnostic plot types
as several figures in the main text. Unless stated otherwise, emitting points
and their corresponding surface footpoints are colored according to footpoint
colatitude in the spin frame, while color saturation indicates emission altitude.

The recurring diagnostic plot types are summarized below:

\begin{description}

\item[Radio emission skymap.]
Total radio emission intensity as a function of spin longitude $\phi$ and observer
colatitude $\thobs$. The intensity is saturated at the 99th percentile of the pixel values
and displayed on a linear scale between zero and the saturation value. This differs from
Fig.~\ref{fig:skymap_PA} and Fig.~\ref{fig:skymap_metric}, where a much larger range of intensity
variation was mapped onto a relatively small range of brightness. The horizontal blue line
marks the selected $\thobs$.

\item[OFLR surface map.]
Open field line regions are shown in gray on the pale blue stellar surface.
Footpoints of field lines contributing to emission at the selected $\thobs$ are
overplotted. The horizontal blue line marks $\thobs$.
Where present, gray curves mark the boundaries of the corresponding pure dipole
OFLRs for the same $\thD$.

\item[3D magnetospheric view.]
Emitting field lines and emission points are shown from selected viewing directions.
When three views are present, the middle panel approximately corresponds to a view of the
XZ plane, while the left and right panels show the YZ plane from two opposite directions.
Last open field lines for the corresponding $P$ are shown in gray for reference.

\item[PA plot.]
Numerically calculated PA values are shown as a function of spin longitude $\phi$.
The analytical RVM prediction is shown by the dashed curve.

\end{description}

\begin{figure*}
\centering
 \includegraphics[width=\textwidth, clip]{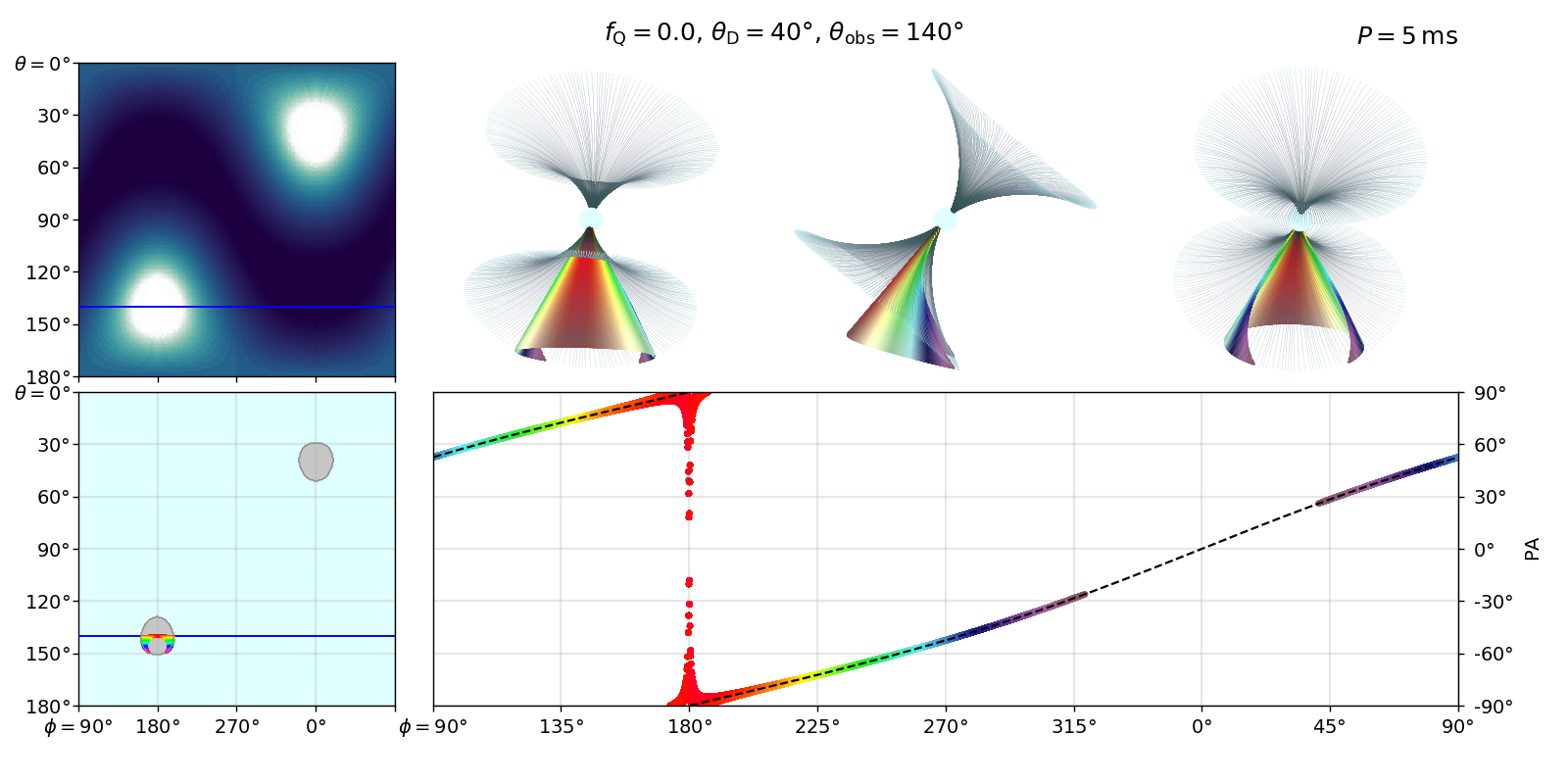}
   \caption{Example of a pure dipole field for the case $\thobs=180\degree-\thD$. Close to $\phi=180\degree$, the 
   PA depends sensitively on the exact orientation of the field line tangent, producing a large spread of PA 
   values around the analytical RVM curve. The \textit{upper left} panel shows the radio emission skymap, the \textit{lower 
   left} panel shows the OFLR surface map, the \textit{upper right} panels show the 3D magnetospheric views, 
   and the \textit{lower right} panel shows the PA plot. The panel content, color coding, and plotting 
   conventions are described in the Appendix glossary.
 }
\label{fig:PA_scenario_D}
\end{figure*}

\begin{figure*}
\centering
 \includegraphics[width=\textwidth, clip]{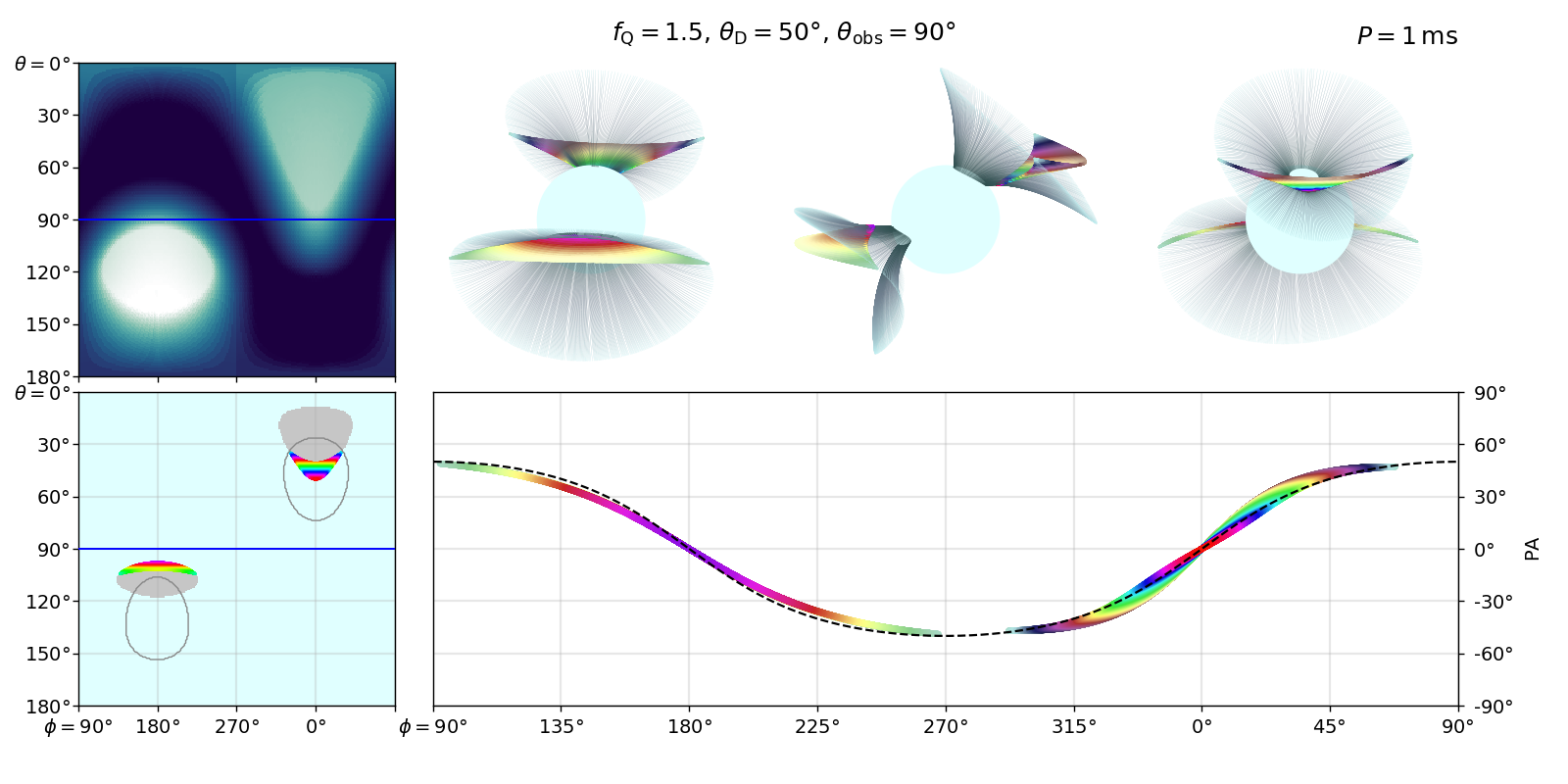}
   \caption{Representative example of a PA track that remains dipole-like despite the presence of a 
   quadrupole component. The PA is largely independent of emission altitude and closely follows the 
   pure dipole RVM prediction, even though the underlying OFLR morphology and emitting field lines 
   differ substantially from those of a pure dipole. The plot layout and visual conventions are the 
   same as in Fig.~\ref{fig:PA_scenario_D} and are described in the Appendix glossary.
}
\label{fig:PA_scenario_A}
\end{figure*}
\begin{figure*}
\centering
 \includegraphics[width=\textwidth, clip]{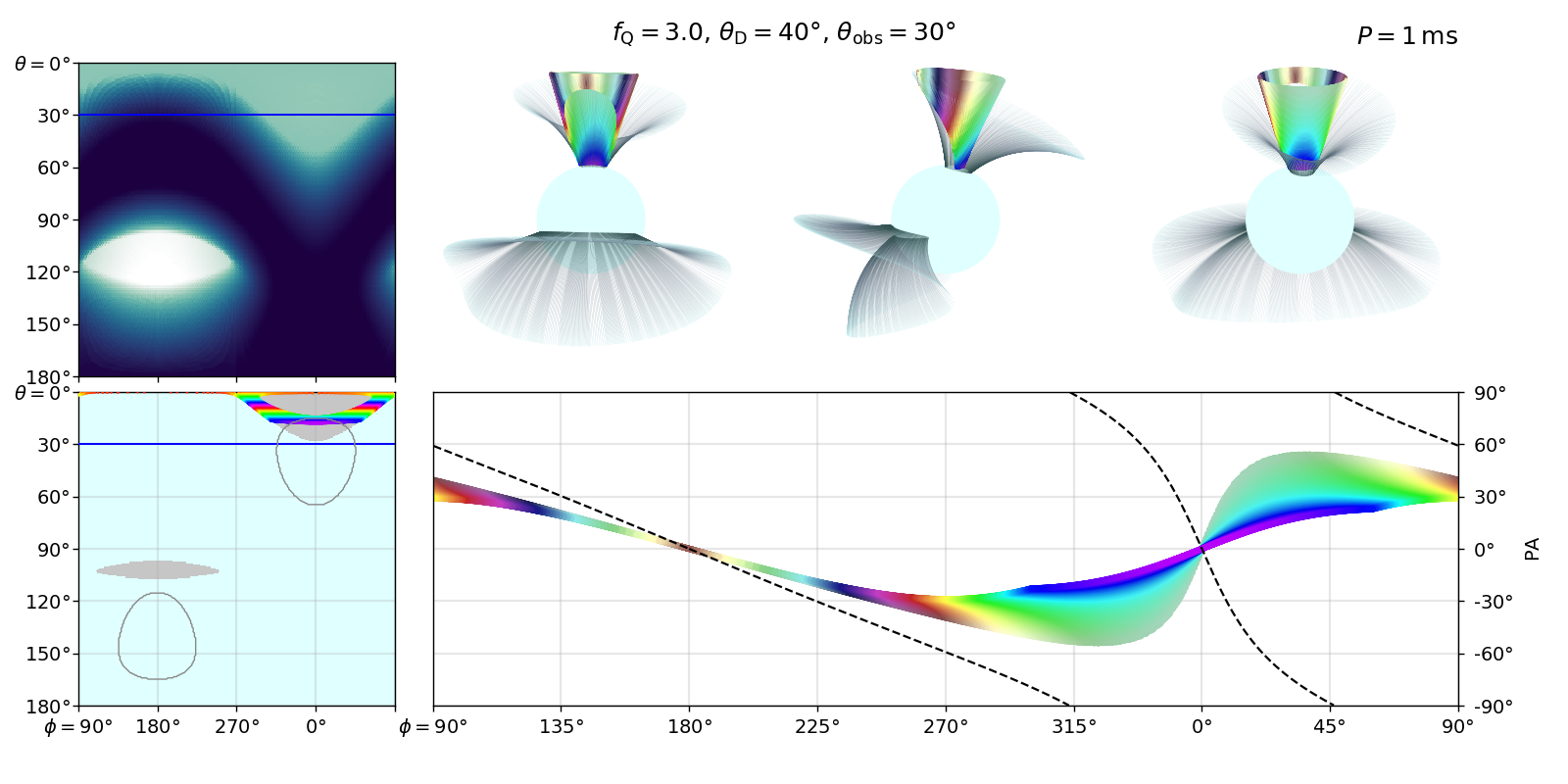}
   \caption{Representative example of the tightest PA curves in our explored parameter space 
   that differ substantially from the RVM prediction for the dipolar component alone. 
   The plot layout and visual conventions are the same as in Fig.~\ref{fig:PA_scenario_D} and are described in the Appendix glossary.}
\label{fig:PA_scenario_E}
\end{figure*}

 \begin{figure*}
\centering
\includegraphics[width=0.45\textwidth, clip]{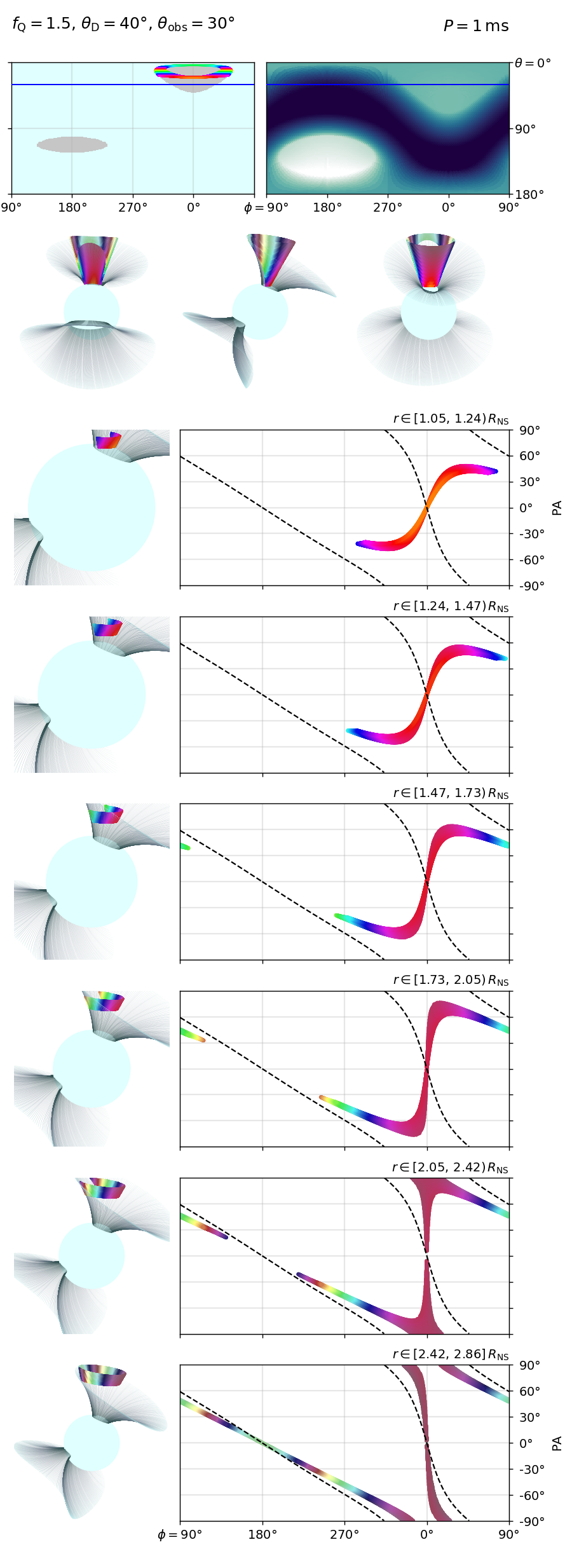}\includegraphics[width=0.45\textwidth, clip]{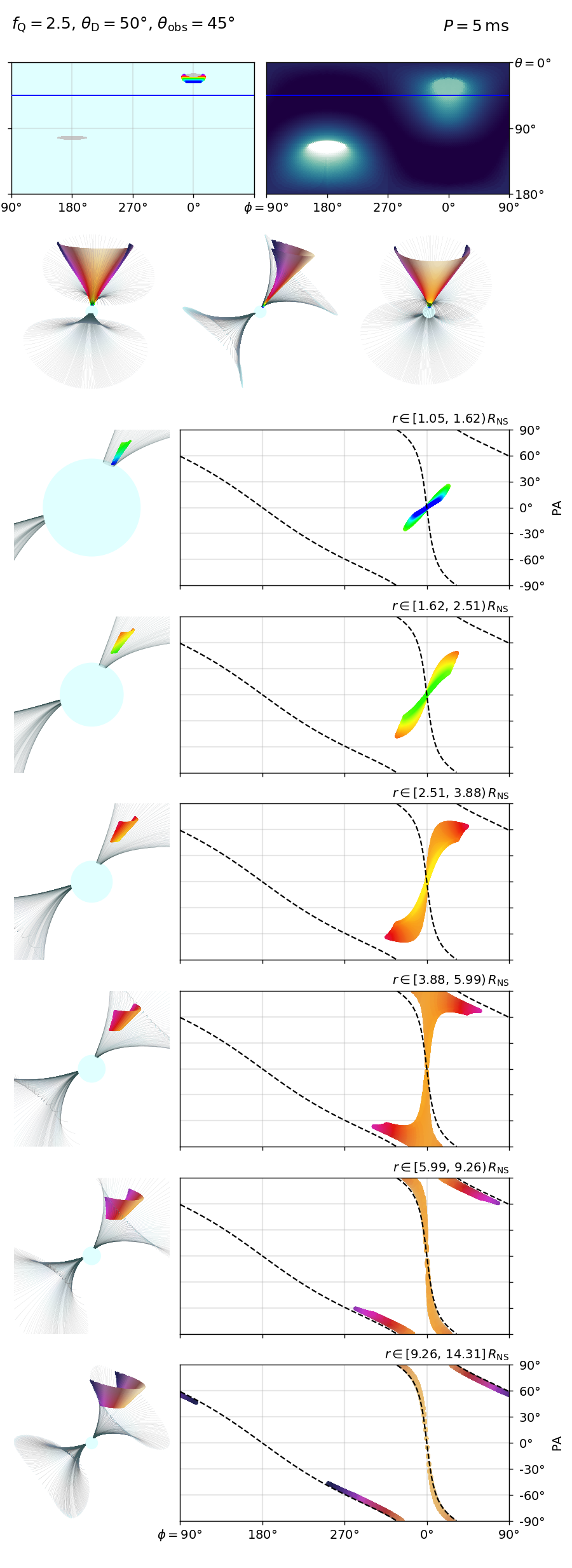}
\caption{Representative examples of a scenario in which the PA varies gradually with emission altitude 
while forming relatively tight RVM-like curves within narrow altitude bands, shown for \Po\ (left)
and \Pf\ (right).
\textit{Top}: OFLR footprints, radio emission skymaps, and 3D views of emitting 
points for the selected $\thobs$.
\textit{Lower panels}: Emitting points and corresponding PA distributions separated into radial 
intervals, indicated above each PA panel.
To better resolve the evolution of PA with altitude, the altitude bands are spaced logarithmically.}
\label{fig:PA_scenario_B}
\end{figure*}

\begin{figure*}
\centering
\includegraphics[width=0.45\textwidth, clip]{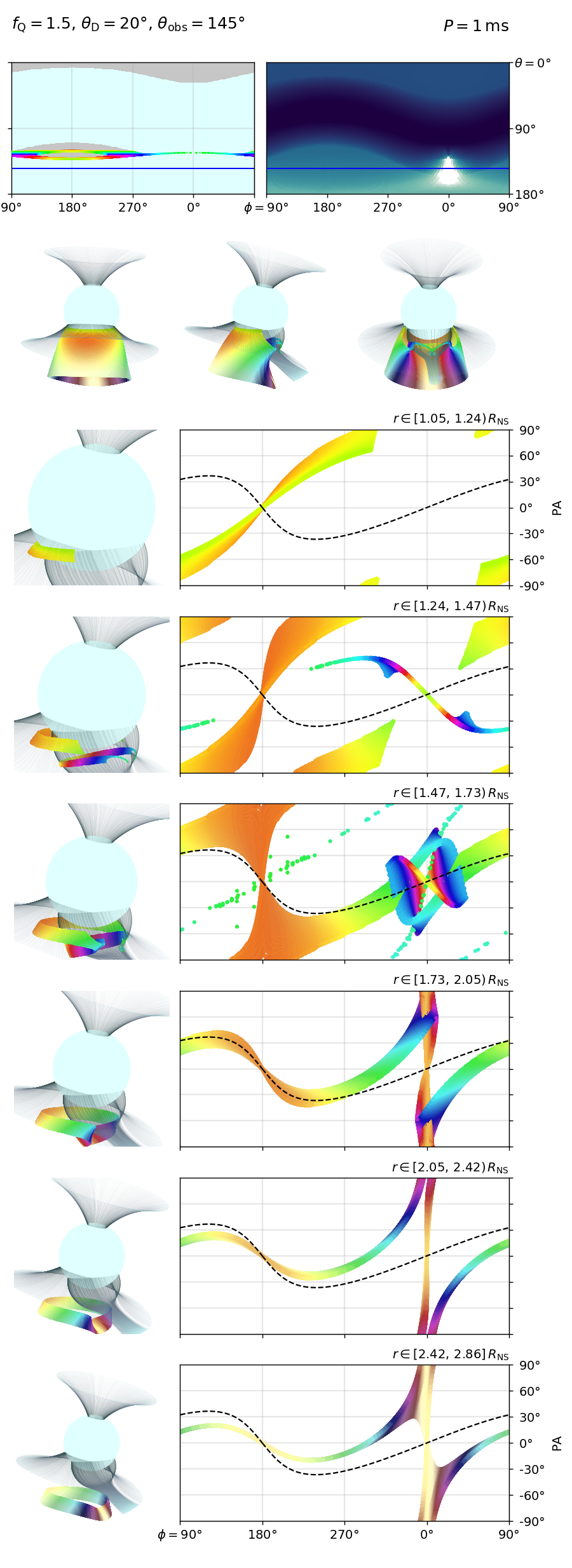}\includegraphics[width=0.45\textwidth, clip]{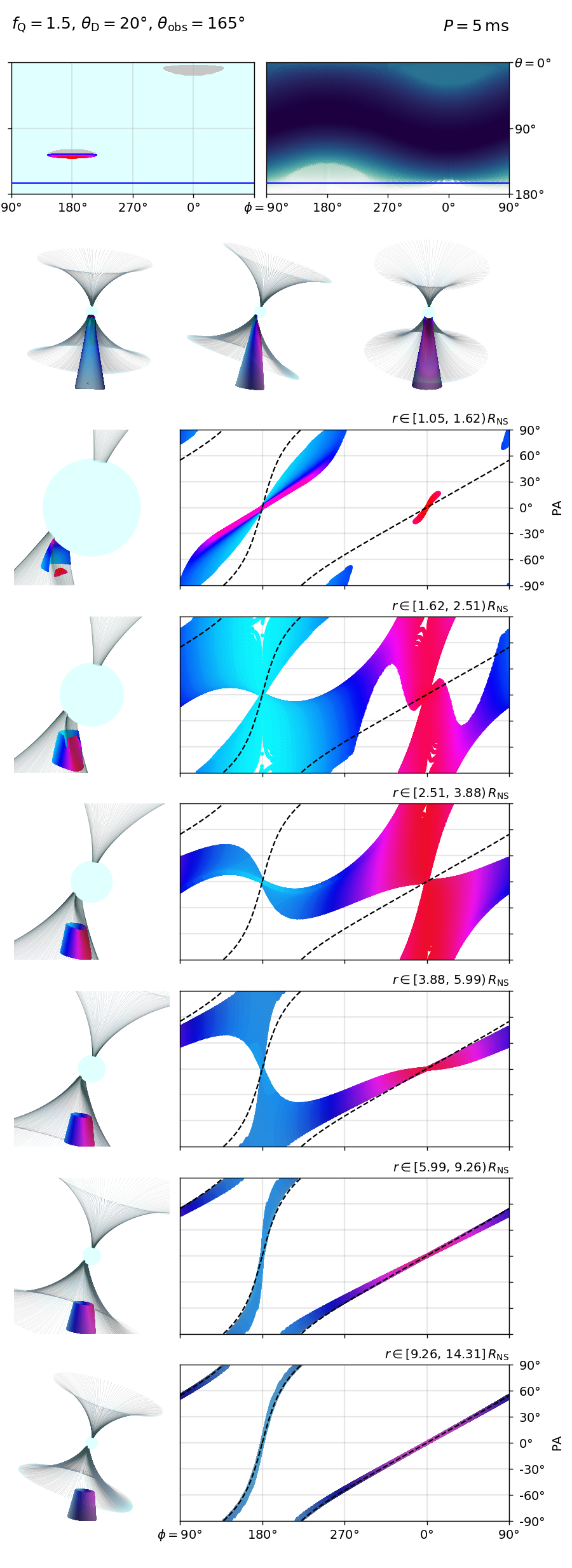}
\caption{Representative examples of a scenario where PA depends strongly on emission altitude and on the 
region of the OFLR contributing to the emission, producing multiple distinct or overlapping PA tracks. 
The case for small $\thD$ is shown for both \Po\ (left) and \Pf\ (right).
Plotting conventions are the same as in Fig.~\ref{fig:PA_scenario_B}.
}
\label{fig:PA_scenario_C_Psmall}
\end{figure*}

\begin{figure*}
\centering
\includegraphics[width=0.45\textwidth, clip]{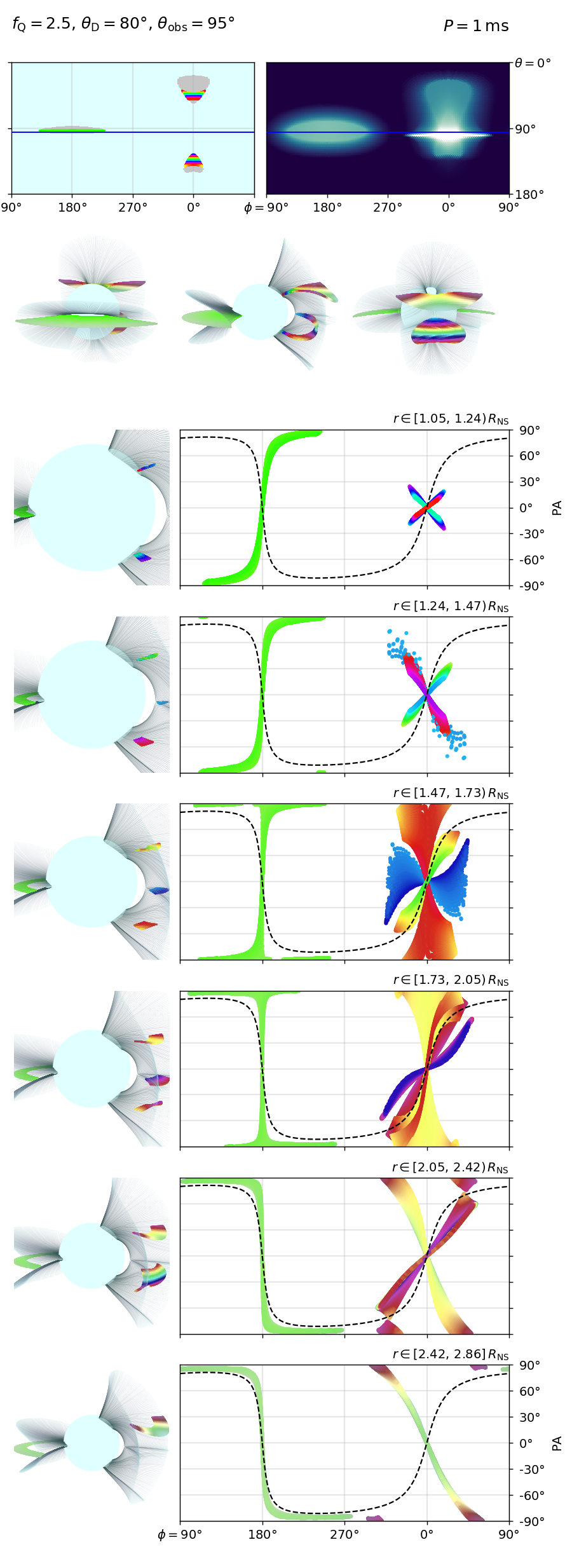}\includegraphics[width=0.45\textwidth, clip]{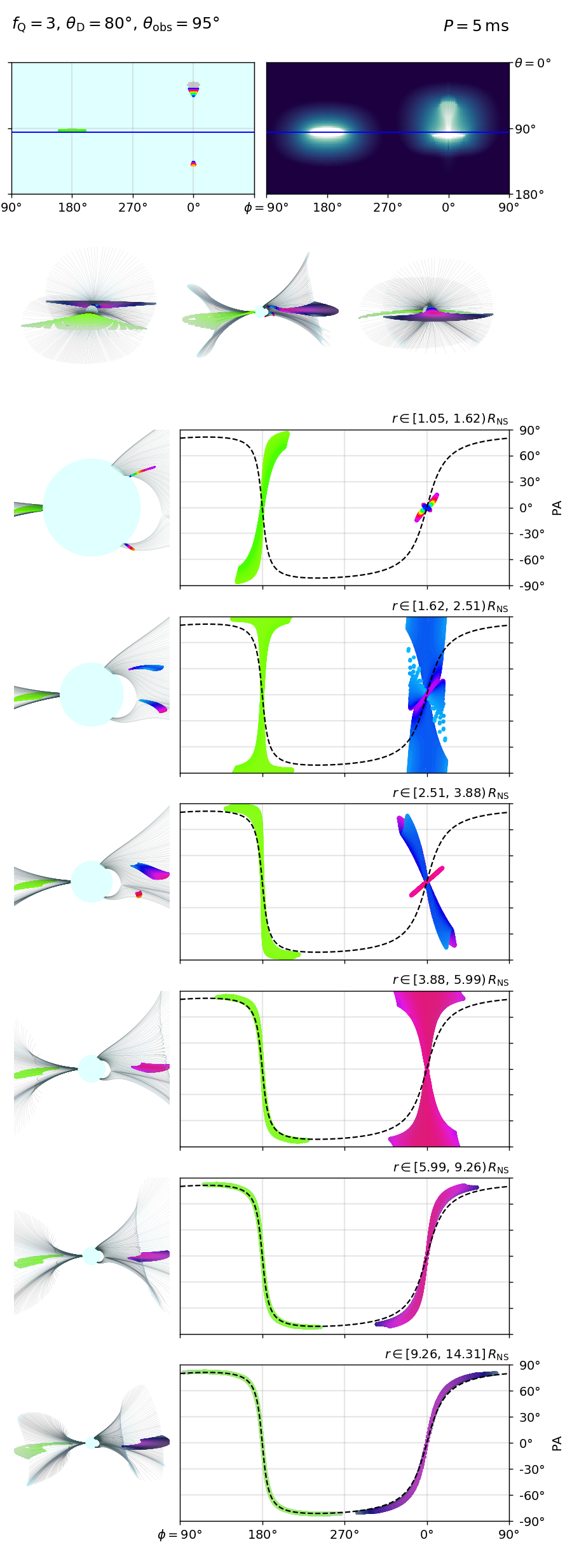}
\caption{Same as Fig.~\ref{fig:PA_scenario_C_Psmall}, but for large $\thD$. }
\label{fig:PA_scenario_C_Pbig}
\end{figure*}

\end{appendix}

\bibliographystyle{aa} 
\bibliography{Multipole_RVM} 
\end{document}

%% file: macros.tex
\newcommand{\blindFitNHeightChunks}{10}

\newcommand{\blindFitRmsPlotMax}{0.5}
\newcommand{\blindFitPhiBinFracMin}{0.1}

\newcommand{\PillA}{$P=0.5$\,ms}
\newcommand{\PillB}{$P=1$\,ms}
\newcommand{\PillC}{$P=5$\,ms}
\newcommand{\thDillustr}{10\degree$, $50\degree$, and $90\degree}
\newcommand{\fQillustr}{0$, $1$, $2$, and $3}
\newcommand{\thDedgetype}{10\degree}
\newcommand{\fQedgetype}{1.5}
\newcommand{\RLCedgemarks}{1.5\RNS$, $6\RNS$, and $20\RNS}
\newcommand{\paPhiBinWidth}{5\degree}
\newcommand{\paMetricRhoGZero}{90\degree}
\newcommand{\paMetricPhiRMax}{10\degree}
\newcommand{\paMetricPhiBMax}{80\degree}
\newcommand{\thDmin}{10\degree}
\newcommand{\thDmax}{90\degree}
\newcommand{\thDstep}{10\degree}
\newcommand{\BQmin}{0.0}
\newcommand{\BQmax}{3.0}
\newcommand{\BQstep}{0.5}
\newcommand{\Rmax}{75}
\newcommand{\dt}{0.05}
\newcommand{\fieldBackend}{numba}
\newcommand{\solverBackend}{cyrk}
\newcommand{\Pmin}{0.9}
\newcommand{\RLCedgeO}{$4.77\RNS$}
\newcommand{\RLCedgeF}{$23.85\RNS$}
\newcommand{\rTrimCoeff}{0.6}
\newcommand{\thetaObsTol}{0.2}
\newcommand{\thobsMin}{5\degree}
\newcommand{\thobsMax}{175\degree}
\newcommand{\thobsStep}{5\degree}
\newcommand{\rMinEmiss}{1.05}

\newcommand{\nsideEmissionMap}{32}
\newcommand{\nsideInitial}{256}
\newcommand{\nsideRefined}{1024}
\newcommand{\nPixInitial}{786432}
\newcommand{\pixAreaDegSqInitial}{0.05}
\newcommand{\nsideRefinementFactor}{16}
\newcommand{\angResDegEmissionMap}{1.8}